\documentclass[structabstract]{aa}  
\usepackage{graphicx}
\usepackage{txfonts}

\begin{document}

 \title{Analysis of combined radial velocities and activity of BD+20~1790: evidence supporting the existence of a planetary companion
\thanks{See Acknowledgements},
\thanks{Tables of the photometry and of the RV are only
available at the CDS. }}

   \author{M. Hern\'an-Obispo
          \inst{1}
          \and M. Tuomi
          \inst{2,3}
          \and M.C. G\'alvez-Ortiz
          \inst{4, 2}
          \and A. Golovin
          \inst{5}
          \and J. R. Barnes
          \inst{6}
          \and H. R. A. Jones
           \inst {2}
          \and S. R. Kane
          \inst{7}
          \and D. Pinfield
          \inst{2}
          \and J. S. Jenkins
          \inst{8}
          \and P. Petit
          \inst{9,10}
          \and G. Anglada-Escud\'e
          \inst{11}
          \and S. C. Marsden
          \inst{12}
           \and S. Catal\'an
           \inst{13}           
          \and S. V. Jeffers
          \inst{11}
           \and E. de Castro
          \inst{1}
          \and M. Cornide 
          \inst{1}
          \and A. Garc\'es
          \inst {14}       
           \and M. I. Jones
          \inst{15}
          \and N. Gorlova
          \inst {16}
          \and M. Andreev
          \inst {17}
          }

  \institute{Dpto. de Astrof\'isica y Ciencias de la
  Atm\'osfera, Facultad de F\'isica, Universidad Complutense
  de Madrid, Avda. Complutense s/n, E-28040, Madrid, Spain 
              \\
              \email{mhernanobispo@fis.ucm.es}
         \and
          Centre for Astrophysics Research, University of Hertfordshire, College Lane, Hatfield, Hertfordshire AL10 9AB, UK
         \and
         University of Turku, Tuorla Observatory, Deparment of Physics and Astronomy, V\"ais\"al\"antie 20, Fl-21500, Piikki\"o, Finland
         \and
         Centro de Astrobiolog\'ia (CSIC-INTA), Ctra. Ajalvir km 4, E-28850 Torrej\'on de Ardoz, Madrid, Spain
         \and 
         Main Astronomical Observatory of  National Academy of Sciences of Ukraine, Zabolotnogo str., 27, Kiev, 03680, Ukraine
         \and
         Department of Physical Sciences, The Open University, Walton Hall, Milton Keynes, MK7 6AA, U. K.
           \and
         Department of Physics \& Astronomy, San Francisco State University, 1600 Holloway Avenue, San Francisco, CA 94132, USA
          \and 
          Departmento de Astronom\'ia, Universidad de Chile, Camino del Observatorio 1515, Las Condes, Santiago, Chile, Casilla 36-D
          \and
           Universit\'e de Toulouse, UPS-OMP, Institut de Recherche en Astrophysique et Plan\'etologie, Toulouse, France
            \and
            CNRS, Institut de Recherche en Astrophysique et Plan\'etologie, 14 Avenue Edouard Belin, F-31400 Toulouse, France 
            \and
	Universit\"at G\"ottingen, Institut f\"ur Astrophysik, Friedrich-Hund-Platz 1, 37077 G\"ottingen, Germany
	\and
	Computational Engineering and Science Research Centre, University of Southern Queensland, Toowoomba, 4350, Australia
           \and
	Department of Physics, University of Warwick, Coventry CV4 7AL, UK
	\and
          Institut de Ci\' ences de l'Espai (IEEC-CSIC), Facultad de Ci\' encies, Campus UAB, 08193, Bellaterra, Spain
          \and
          Department of Electrical Engineering and Center of Astro-Engineering UC, Pontificia Universidad Cat\'olica de Chile, Av. Vicu\~na Mackenna 4860, 782-0436 Macul, Santiago, Chile
           \and
           Institute of Astronomy, Katholieke Universiteit Leuven, Celestijnenlaan 200D BUS 2401, 3001 Leuven, Belgium
           \and
            Terskol Branch of Institute of Astronomy RAS, Kabardino-Balkaria Republic, 361605 Terskol, Russia
             }

   \date{recieved -- ; accepted --}

 \abstract{ In a previous paper we reported a planetary companion to the young and very active K5Ve star BD+20 1790. We found that this star has a high level of stellar activity ($log R^{\prime}_{HK}$=-3.7) that manifests in a plethora of phenomena (starspots, prominences, plages, large flares). Based on a careful study of these activity features and a deep discussion and analysis of the effects of the stellar activity on the radial velocity measurements, we demonstrated that the presence of a planet provided the best explanation for the radial velocity variations and all the peculiarities of this star. The orbital solution resulted in a close-in massive planet with a period of 7.78 days. However, a paper by Figueira et al. (2010) questioned the evidence for the planetary companion.}
{This paper aims to more rigorously assess the nature of the radial velocity measurements with an expanded dataset and new methods of analysis.}
{We have employed Bayesian methods to simultaneously analyse the radial velocity and activity measurements based on a combined dataset that includes new and previously published observations.}
{We conclude that the Bayesian analysis and the new activity study support the presence of a planetary companion to BD+20 1790. A new orbital solution is presented, after removing the two main contributions of stellar jitter, one that varies with the photometric period (2.8 days) and another that varies with the synodic period of the star-planet system (4.36 days). We present a new method to determine these jitter components, considering them as second and third signals in the system. A discussion on possible star-planet-interaction is included, based on the Bayesian analysis of the activity indices, which indicates that they modulate with the synodic period. We propose two different sources for flare events in this system: one related to the geometry of the system and the relative movement of the star and planet, and a second one purely stochastic source that is related to the evolution of stellar surface active regions. Also, we observe for the first time the magnetic field of the star, from spectropolarimetric data.}{}

\titlerunning{Bayesian analysis of RV and activity for BD+20 1790}

   \maketitle
%

\section{Introduction}
The quest for planets around other stars has become one of the most successful and productive fields in Astronomy. Since the surprising discovery of 51 Peg b by Mayor \& Queloz in 1995, the ever increasing rate of newly reported planets \footnote{exoplanet.eu} has revealed a large variety of new and strange worlds, in some cases very different from the ones in our Solar System.

The exoplanetary zoo includes a wide range of properties and orbital configurations even challenging the theories of planetary formation and evolution (Udry \& Santos 2007; Mordasini et al. 2009a, b). One of the most unexpected varieties of these ``exo-worlds" were massive planets in close short orbits around their host stars, the so-called {\it hot Jupiters}. Detection techniques such as the radial velocity (RV) method are especially sensitive to these massive exoplanets. Unfortunately, they are also sensitive to the effect of stellar activity. The imprint of a spot in the RV variations could mimic a planetary reflex motion with a period near to the stellar rotational period (Saar \& Donahue 1997, Desort et al. 2007).     
     This effect is more pronounced when studying young and active stars, in which the magnetic field and the phenomena related to it seem to occur on a more enhanced scale (see e.g. Hall 2008). The contamination by stellar activity could swallow the signal from a potential planetary companion and also provide false planetary detections (Queloz et al. 2001, Hu\'elamo et al. 2008). Hence, a detailed study and characterization of stellar activity, to achieve a deeper understanding of its effects, are critical for disentangling the origin of the RV variations.

The analysis of observations based on the law of conditional probabilities, i.e. Bayes' rule, has become one of the most powerful tools in extracting weak signals of planetary companions from noisy data (e.g. Gregory 2007a, b, Tuomi \& Kotiranta 2009, Tuomi et al. 2011, Tuomi 2012, Tuomi et al. 2013, Jenkins et al. 2013, Jenkins \& Tuomi 2014, Tuomi et al. 2014). The method enables us to equip each model with a probability of it being correct with respect to the other models given the observations (e.g. Ford \& Gregory 2007). In addition, it is possible to receive approximations for the probability distributions of model parameters using the various posterior sampling techniques. All of these allow one to determine reliable and robust  results from noisy data.\\

In this paper we present a simultaneous and complementary Bayesian analysis of the RVs and the chromospheric activity indices of BD+20 1790. Both are measured from the same spectra. The goal is to check whether the signal of the planetary companion proposed is still detected or is an artefact (artificial signal due to the sampling of the observations). 

We model the measurement noise as in Tuomi et al. (2011, 2013, 2014), and do not make any assumptions regarding its magnitude in the different datasets from the various telescope-instrument combinations.
As pointed out by Saar (2009) the stellar jitter is the sum of several components, caused by the different activity features (spots, plages, flares, ...), stellar convection, etc. In addition to all these possible jitter sources, the stellar magnetic field configuration evolves and changes with time. In the "classic way" the stellar jitter is fixed and determined from empirical equations that might underestimate or overestimate it, and consider only one source for the stellar jitter, mainly summarized with the log $R^{\prime}_{HK}$ index. We present a new way to determine the stellar jitter without making any assumption about the jitter values. We determine that the RV variations come from a combination of three signals, where the main signal corresponds to the component from the planetary companion and the two others are related to activity.\\

The paper is organized as follows. A review of stellar and planetary companion properties and our previous work is summarised in Sect. 2. New photometric, spectroscopic and spectropolarimetric observations are shown in Sect.~3. The results of the combined Bayesian analysis of RV and activity indices are provided in Sect.~4. The new orbital solution for BD+20 1790 system is presented in Sect.~5. Results are presented in the context of a star-planet interaction scenario in Sect.~6. Finally, our results are discussed in Sect.~7 and summarized in Sect.~8.

\section {BD+20 1790 system: previous works}

BD+20 1790 is a young exceptionally active K5Ve BY Dra late-type star. Main properties are detailed at Table~\ref{starplanet}. In our previous studies (Hern\'an-Obispo et al. 2005, 2007, 2010, L\'opez-Santiago et al. 2006, 2010), BD+20 1790 shows a high level of stellar activity that is manifested in a plethora of phenomena at different atmospheric layers. 

\begin{figure}
  \centering
\includegraphics [scale=0.30]{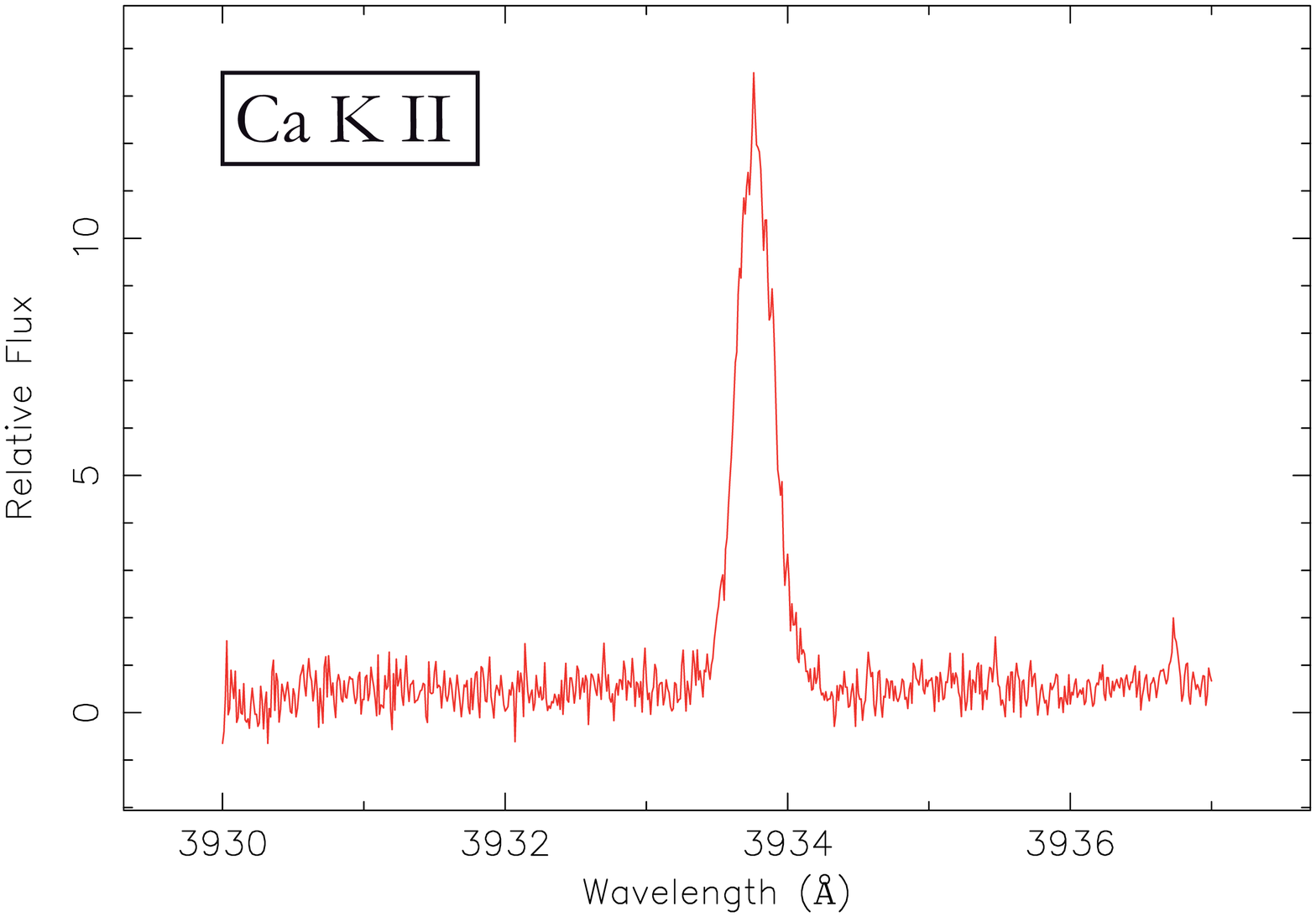}
\includegraphics [scale=0.30]{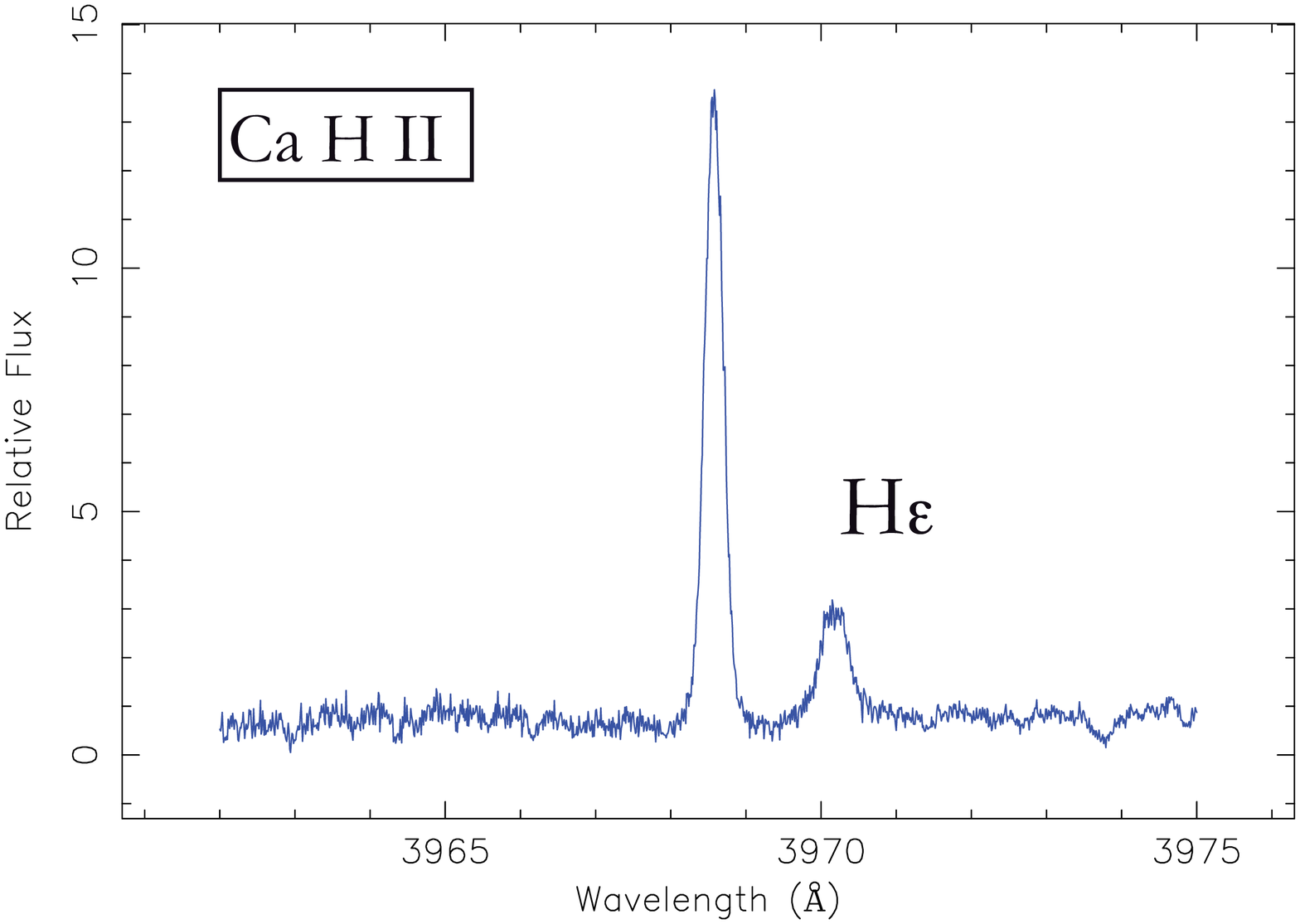}
       \caption{An example of Ca~{\sc ii} H \& K lines for BD+20 1790 at quiescent state. Both lines appear always in strong emission over continuum, and do not show the reversal emission observed in other active stars.}
        \label{fig:hyk}
\end{figure}

We detected strong chromospheric emission, as suggested by all the chromospheric activity indicators being in emission above continuum (from Ca~{\sc ii} H \& K to Ca~{\sc ii}  infrared triplet). The Ca~{\sc ii} H \& K lines do not show clear evidence of a strong absorption feature with an emission reversal core, as seen in other active stars. Instead these lines always appeared in emission, indicating strong chromospheric heating (see Fig.\ref{fig:hyk}). The average $log R^{\prime}_{HK}$=-3.7, computed considering all the data, with and without flare state, is greater than that for Corot-2A, which is cited as one of the most active stars hosting a planet (Schr{\"o}ter et al. 2011).

Prominence-like structures were detected at latitudes far from the equatorial plane (Hern\'an-Obispo et al. 2005), as well as plage-like structures. From the modulation of the photometric light curve we reported the presence of photospheric spots, with a mean spot filling factor ({\rm $f_s$} hereafter) up to 9\%. We detected large energetic flares (energy released up to 10$^{37}$ erg) with a temporal decay of more than 3 hours, and a high rate of occurrence. The X-ray luminosity of BD+20 1790, $logL_X$=29.2, is high if we take into account that for K stars of similar youth the estimated $logL_X$ is about 28.8. This excess of X-ray emission could be related to large flares.

In comparison with objects with the same spectral type, rotational velocity ($v{\rm sin}i$) and age range, BD+20 1790 presents a very high level of stellar activity. We compare the activity of this star with a large catalogue of single and binary stars collected and characterized by Montes et al (2001), L\'opez-Santiago et al. (2006), L\'opez-Santiago et al. (2010), L\'opez-Santiago (2005), G\'alvez-Ortiz (2005). BD+20 1790 has a level of activity greater even than the binary systems in the catalogue. 

The overall picture shows that the magnetic field of this star is a complicated puzzle to piece together.\\

L\'opez-Santiago et al. (2006) proposed its membership to the AB Dor moving group, with an estimated age-range between 35-80 Myr, using the equivalent width of Li I~6707~{\AA}. We note here that there is a parallel debate in the literature about the age of the AB Dor moving group and the AB Doradus quadruple system itself. Age estimations range from 30 - 70 Myr, comparable with the youngest known open clusters (Zukerman et al. 2004, Lopez-Santiago et al. 2006, Close et al. 2007); to 150 Myr, coeval with the Pleiades open cluser (Luhnman et al. 2005, Ortega et al. 2007, da Silva et al. 2009, Barenfeld et al. 2013). For the AB Doradus system itself ages proposed ranged from 40 - 50 Myr (Guirado et al. 2011) to 50-100 Myr (Janson et al. 2007). 

From Mamajek \& Hillebrand (2008; MH08), by using the fractional X-Ray luminosity $R_X$, we inferred an age of up to 35 Myr for BD+20 1790. Also, from MH08 Eq [3], age can be estimated from $log R^{\prime}_{HK}$. Considering a mean value at quiescent state for $log R^{\prime}_{HK}$ of -4.1, we determined an age of up to 58 Myr. An additional estimation can be obtained by using gyrochronology. From the MH08 relations, that improved the ones in Barnes (2007), we derived an age of up to 38 Myr.\\

From the analysis of 91 RV data points taken in different epochs spanning six years, a careful study and deep discussion about the effects of the stellar activity on the RV measurements, we proposed in Hern\'an-Obispo et al. (2010) (hereafter Paper I) the existence of a planetary companion to explain both the RV variations and the high stellar activity level. Readers are referred to Sect.~4 in Paper I for further details on that study and the discussion on the nature of RV variations. The Keplerian fit of the RV data yielded a solution for a close-in massive planet, with a period of 7.78 days and a $M \sin i$ $\approx$ 6 M $_J$.\\

\begin{table}
\caption{Parameters of BD+20 1790 and the planet candidate}             
\label{starplanet}      
\centering                          
\begin{tabular}{l r r}        
\hline\hline                 
Parameter & Value & Reference  \\    
\hline                        
            Spectral Type & K5 V & (a)\\
            $B-V$ & 1.15 & \\
            $M (M_{\sun})$ & 0.63 $\pm$ 0.09 & (b)\\
            $T_{\rm eff}$ & 4410  K & (b, c)\\
            $\log g$ & 4.53 $\pm$ 0.17 & (b)\\
            $EW{\rm(Li)}$ & 110 $\pm$ 3  m\AA & (b)\\
            $Distance$ & 25.4 $\pm$ 4 $pc$ & (d)\\
            $Age$ & 35 - 80  Myr & (e)\\
            $v{\rm sin}i$ & 10.03 $\pm$ 0.47  km s$^{-1}$ & (b, f)\\
            $P_{\rm phot}$ & 2.801 $\pm$ 0.001  days & (b)\\
            $i$ & 50.41  degrees& (b)\\
            $R$ & 0.71 $\pm$ 0.03 $R_{\sun}$ & (b)\\
            $[Fe/H]$ & 0.30 $\pm$ 0.20 & (b)\\
            $L_X$ & 1.6 $\pm$ 0.5 $10^{29}$ erg s$^{-1}$ & (b) \\
            $L (L_{\sun})$ & 0.17 $\pm$ 0.04  & (b)\\
            $log R^{\prime}_{HK}$ & -3.7 & (g)\\
  
\hline\hline   

\end{tabular}
\begin{list}{}{}
\item (a) Jeffries (1995); (b) Hern\'an-Obispo et al. (2010); (c) McCarthy \& White (2012); (d) Reid et al. (2004); (e) L\'opez-Santiago et al. (2006); (f) L\'opez-Santiago et al. (2010); (g) This work.
\end{list}
   \end{table}

Figueira et al. (2010) (hereafter F10) questioned the planet candidate, providing new RV data obtained with CORALIE, that span 55 days between December 2009 to February 2010. They carried out two campaigns (of 21 and 8 days respectively); finding RV amplitudes significantly different from those reported in Paper I, up to 230 m s$^{-1}$ peak-to-peak for Set 1 (December) and up to 460 m s$^{-1}$ for Set 2 (February). This difference in RV amplitude between the two sets of CORALIE data (second is twice the first set) and with the RV amplitude derived in Paper I (almost 900 m s$^{-1}$), is intriguing. The CORALIE radial velocity data suggest a dramatic decrease of the strength of the magnetic field. As it is presented in forthcoming sections, twelve years of photometric and spectroscopic monitoring (including contemporaneous with the F10 data) indicate to us that the stellar activity never reaches the debilitation levels that the CORALIE radial velocities imply. Although there are variations between seasons, the lowest level of activity is greater than the CORALIE radial velocities suggest. Conversely, the analysis of the activity indicators of CORALIE data shows that the star has a high level of stellar activity.

\begin{table*}
\caption[]{Observing runs
\label{runs}}
\begin{center}
\small
\begin{tabular}{lllllllll}
\noalign{\smallskip}
\hline  \hline
\noalign{\smallskip}
 Date & Telescope & Instrument & CCD chip & Spect. range & Orders &
Dispersion & FWHM  & S/N\\
   &           &            &   \#     & ~~~~~~~(\AA)   &          & ~~~~~(\AA/pix) & ~~~(\AA) & \\
\noalign{\smallskip}
\hline
\noalign{\smallskip}
31/01/2008 & 2.2m$^{\rm a}$  & FOCES & 2048x2048 15$\mu$m LORAL$\#$11i & 3830 - 10850 & 96 & 0.03 - 0.07 & 0.09 - 0.26 & 80\\
31/03-2/04/2010 & TNG $^{\rm b}$  & SARG & 2148x4200 13.5$\mu$m EEV & 5480 - 10120 & 51 & 0.08 - 0.14 & 0.16- 0.32 & 140\\
9-11/02/2009 & NOT$^{\rm c}$  & FIES & 2000x2000 15$\mu$m EEV42-40 & 3620 - 7360 & 79 & 0.02 - 0.04 & 0.05 - 0.11 & 70\\
22-25/12/2010 & MERCATOR $^{\rm d}$  & HERMES & 2048x4608 24$\mu$m E2V42-90 & 3770 - 9000 & 58 & 0.016 - 0.13 & 0.08 - 0.35 & 120 \\
27-31/12/2012 & TNG $^{\rm b}$  & HARPS-N & 4k4 24$\mu$m E2V & 3830 - 6930 & 68 & 0.03 - 0.06 & 0.03 - 0.06 & 30 \\
22-25/02/2013 & TNG $^{\rm b}$  & HARPS-N & 4k4 24$\mu$m E2V & 3830 - 6930 & 68 & 0.03 - 0.06 & 0.03 - 0.06 & 30 \\
21-24/01/2014 & TNG $^{\rm b}$  & HARPS-N & 4k4 24$\mu$m E2V & 3830 - 6930 & 68 & 0.03 - 0.06 & 0.03 - 0.06 & 30 \\
18/05/2014 & TNG $^{\rm b}$  & HARPS-N & 4k4 24$\mu$m E2V & 3830 - 6930 & 68 & 0.03 - 0.06 & 0.03 - 0.06 & 30 \\
\noalign{\smallskip}
\hline \hline
\noalign{\smallskip}
\end{tabular}
\end{center}
\vspace{-0.25cm}
{\small
$^{\rm a}$ 2.2~m telescope at the German Spanish Astronomical Observatory (CAHA)
(Almer\'{\i}a, Spain).\\
$^{\rm b}$ 3.58~m {\it Telescopio
  Nazionale Galileo} (TNG) at Observatorio del Roque de los Muchachos
(La Palma, Spain).\\
$^{\rm c}$ 2.5~m {\it Nordic Optical Telescope} (NOT) at Observatorio del Roque de los Muchachos
(La Palma, Spain).\\
$^{\rm d}$ 1.2~m MERCATOR Telescope at Observatorio del Roque de los Muchachos
(La Palma, Spain).\\
}
\end{table*}

\section{Observations and Data Analysis}

In addition to the data reported in Paper I, photometric, spectroscopic and spectropolarimetric observations were taken over 2009-2014, to continue our study and characterization of the stellar activity of this object. Details of the observations are given at Table~\ref{runs}, Table~\ref{asaslog} and Table~\ref{terskollog}. To see details of observations from Paper I, readers are referred to Table 2 at Hern\'an-Obispo et al. (2010).

\subsection{Spectroscopic data}

Our initial observational strategy was designed to spectroscopically follow-up the time varying chromospheric activity indicators, with high temporal and spectral resolution. In addition to these, spectroscopic observations with HARPS-N were carried out, in four separated runs. Although bad weather prevent us to get the expected number of RV points. Apart from further compilation of high quality RV data to discriminate the nature of the RV variations, HARPS-N data were used to compare with the CORALIE data, and to confirm whether we were able to detect the planetary signal. 

Except for the HERMES and HARPS-N data, the rest of the data, target star and standards, were reduced following standard procedures using the routines in the IRAF\footnote{IRAF is distributed by the National Optical Observatory, which is operated by the Association of Universities for Research in  Astronomy, Inc., under contract with the National Science Foundation.} {\it echelle} package. The spectra taken with HERMES (Raskin et al. 2011) were reduced using a dedicated automated data reduction pipeline and radial velocity toolkit (HermesDRS version 4.0) that provides full data reduction and calibration. HARPS-N pipeline supply science quality extracted spectra through a standard process:
 localization of the spectral orders on the 2D-images, optimal order extraction, background subtraction, cosmic-ray rejection, corrections of flat-field and wavelength calibration (Cosentino et al. 2012).\\

\begin{figure}
   \centering
    \includegraphics [scale=0.35]{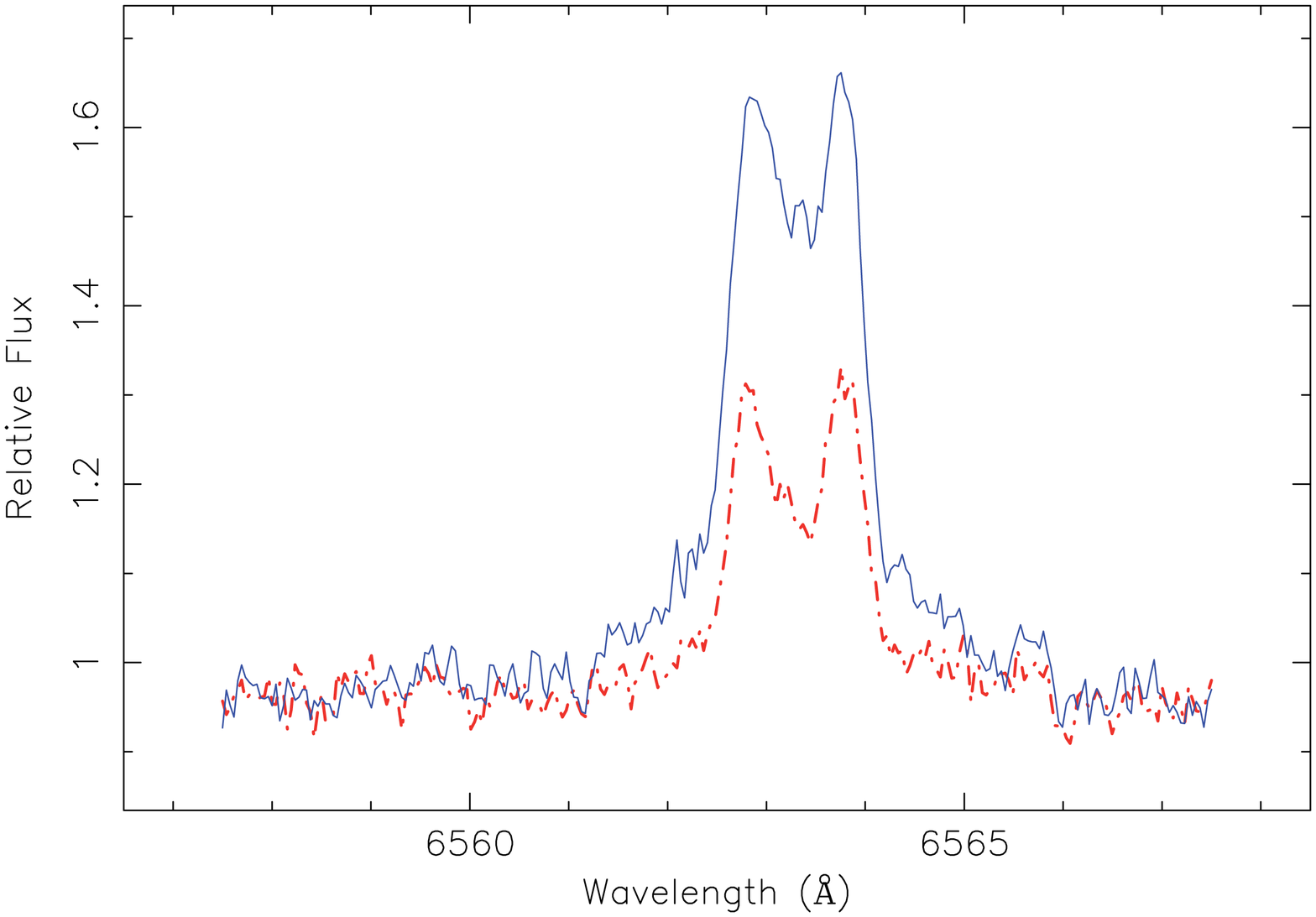}
     \includegraphics [scale=0.35]{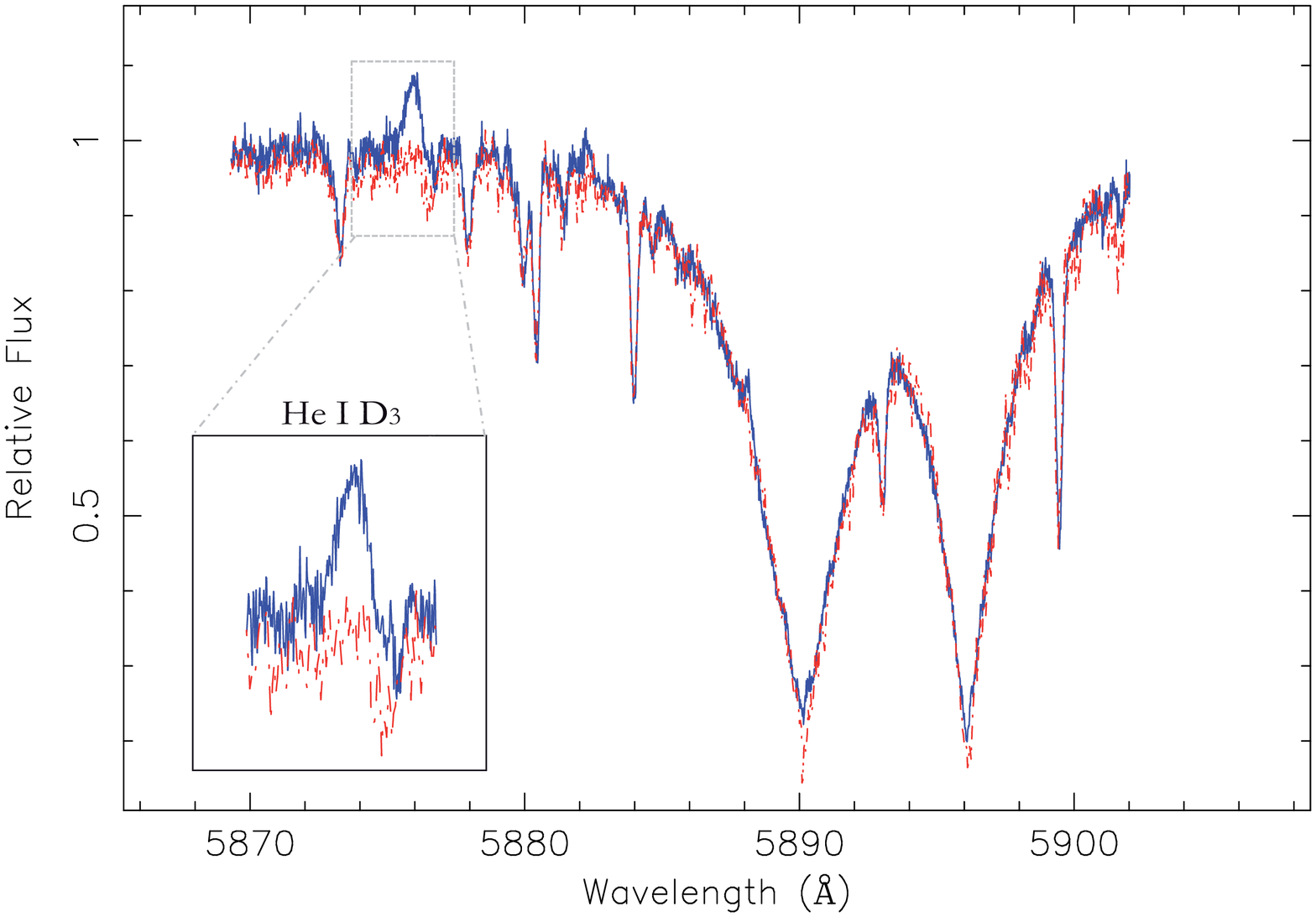}
       \caption{Top: H$\alpha$ line for 2013 February 22 (HARPS-N). Red dashed lines are for quiescent state. The EW (H$\alpha$)  changes between -0.86 \AA~ (quiescent state) to -1.89 \AA~ (flare state). Bottom: He I $D_{3}$, Na I $D_{1}$ and $D_{2}$ lines for 2013 February 22 (HARPS-N). Red dashed lines are for quiescent state. Note the He I $D_{3}$ line (5876 \AA) strong emission and the filling in of the Na I $D_{1}$ and $D_{2}$ lines during the flare.}
               \label{fig:megaflare}
\end{figure}

Heliocentric radial velocities were determined following the same procedure as Paper I (see paper for details). For the spectra taken with HERMES, we measured RV in the same way as the rest of the data, but we also used the RV toolkit from the HERMES pipeline to double check for consistency, which was established. For HARPS-N data, we obtained the RV measurements and corresponding errors through  the HARPS-N on-line pipeline, based on the numerical cross-correlation function  (CCF) method (Baranne et al. 1996) with the weighted and cleaned-mask modification (Pepe et al. 2002), by applying the K5 mask, the same spectral type as BD+20 1790. Moreover, to calculate the mean RV of every spectra we avoided orders that contain emission lines, to especially control the contamination of RV from smaller or moderate flares. As shown by Reiners (2009), giant flares can have a strong effect on RVs. These kinds of strong flare events affected the whole spectrum, not only increasing chromospheric emission lines, but also the continuum including orders far away from emission lines. Reiners (2009) have shown that the RV shift could be up to 600 m/s for large flares on the mid-M star CN Leo, and recommended that these affected data should not be used to determine RV variations (see also Barnes et al. 2014). We detected two large flares in the HARPS-N data, for 2013 February 22 and 2014 January 21. Those flare events are the most powerful we have observed over all runs. In Fig.~\ref{fig:megaflare} is shown the H$\alpha$ and He I $D_{3}$ lines for February 22. The quiescent state is also shown with a dashed line to highlight the increased emission. The He I $D_{3}$ line (5876 \AA) can be seen in absorption in both, Sun and active stars, with the exception of strong solar flares and during stellar flares, where it is observed in emission (Robinson \& Boop 1987, Montes et al. 1997, 1999, Oliveira \& Foing 1999, Garc\'ia-\'Alvarez et al. 2003). He I  emission can also be seen in relation with accretion in really young stars, but no other accretion signatures are seen in the spectra of BD+20 1790.\\

Spectroscopic indices of the main chromospheric activity indicators (Ca~{\sc ii}
H \& K, Balmer lines, He I D$_{3}$ region, Ca~{\sc ii} infrared triplet lines) were measured, following Saar \& Fisher (2000) and Bonfils et al (2007). Due to the different wavelength coverage of the spectrographs, the number of data points is not the same for all the indices. \\

We also analysed the CORALIE data from F10, measuring the activity indices and characterising the chromospheric features and flare events detected. The CORALIE RV-data were provided by Figueira (priv. comm.).

\begin{figure*}
   \centering
\includegraphics [scale=0.50]{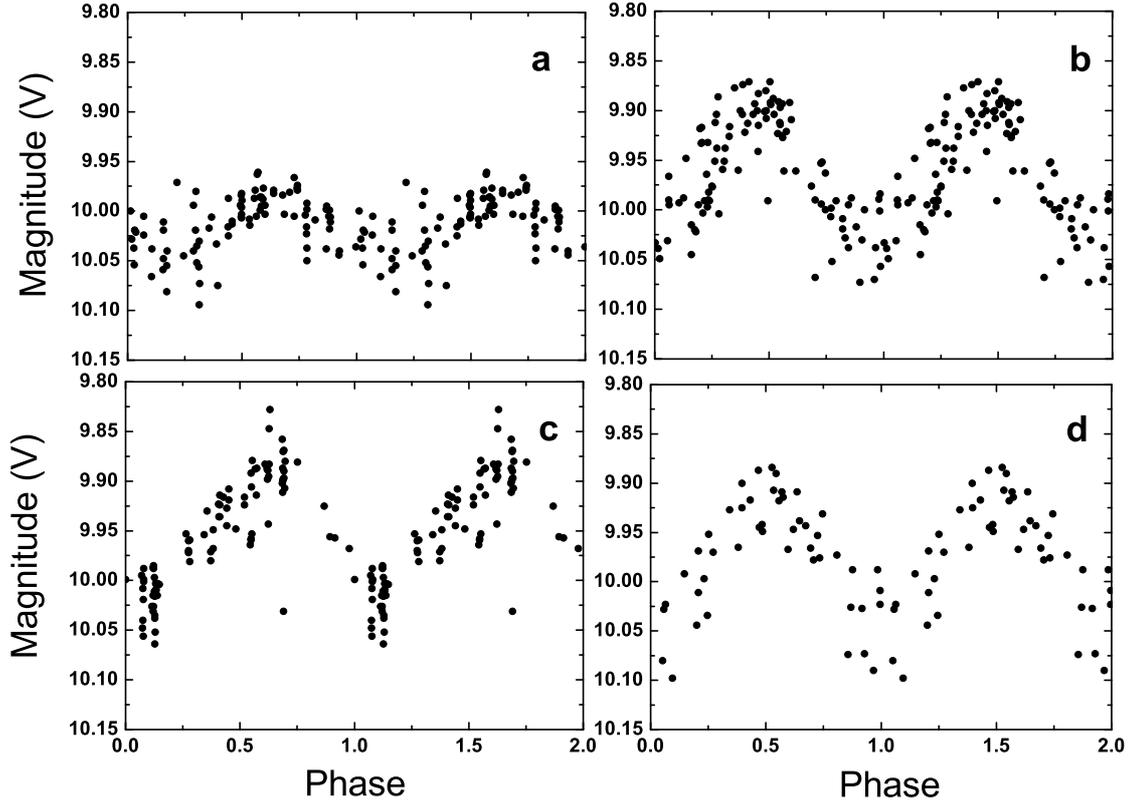}
 \caption{Sample light curves from ASAS data, folded with $P_{rot}$ period.  a) 2002-2003 data; b) 2003-2004 data; c) 2004-2005 data; d) 2007-2008 data; note the scale constancy.}
 \label{fig:phase}
\end{figure*}

\subsection {Photometric data}

\begin{figure}
   \centering
\includegraphics [scale=0.26]{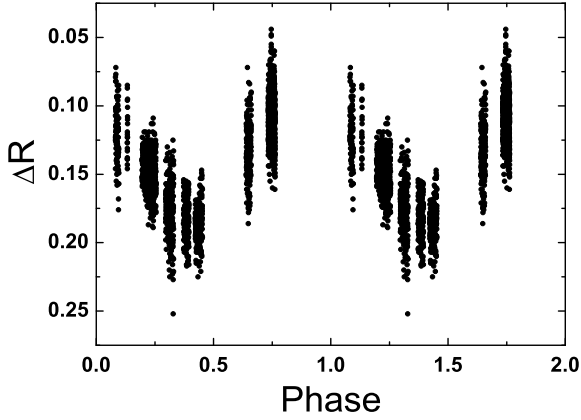}
       \caption{Light curve of BD+20 1790 based on Terskol photometry and folded with rotational period.}
        \label{fig:terskol}
\end{figure}

We were mainly concerned with testing whether the magnetic field could have changed as dramatically as is suggested by the CORALIE radial velocities analysis of F10. To investigate this we performed almost simultaneous spectroscopy and photometry in March-April 2010, even contemporaneous with the CORALIE observations. Photometric follow-up was carried out at the Terskol Observatory (Russia) during 10 nights (see Table~\ref{terskollog}), by using the Zeiss-600 telescope with the PixelVision CCD camera (2 nights) and the Meade 14'' telescope, equipped with STL-1001 CCD camera (8 nights). The data set contains 2816 points. Data were heliocentrically corrected and the light curves were folded with the rotational period as displayed in Fig.~\ref{fig:terskol}.

\begin{table}
\begin{center}
\caption{\label{asaslog} Log of ASAS-observations of BD~+20~1790}
\begin{tabular}{llll}
\hline \hline Year & $T_{start}$  & $T_{end}$ & $N_{points}$ \\
 & 2450000+ & 2450000+ & \\
\hline

2002 - 2003 & 2621 & 2760 & 90 \\
2003 - 2004 & 2912 & 3132 & 105 \\
2004 - 2005 & 3291 & 3425 & 95 \\
2005 - 2006 & 3657 & 3796 & 29 \\
2006 - 2007 & 4089 & 4162 & 10 \\
2007 - 2008 & 4391 & 4591 & 47 \\
2008 - 2009 & 4762 & 4916 & 37 \\

\hline \hline
\end{tabular}
\end{center}
\end{table}

\begin{table}
\begin{center}
\caption{\label{terskollog} Log of Terskol-observations of
BD +20 1790}
\begin{tabular}{lll}
\hline Date & Telescope  & CCD   \\
\hline

26/27 March 2010 &  Z-600 (0.6m) &  PixelVision \\
30/31 March 2010 &  Z-600 (0.6m) &  PixelVision \\
28/29 March 2010 &  Meade (0.35m) & STL-1001 \\
02/03 April 2010 &  Meade (0.35m) & STL-1001 \\
08/09 April 2010 &  Meade (0.35m) & STL-1001 \\
10/11 April 2010 &  Meade (0.35m) & STL-1001 \\
11/12 April 2010 &  Meade (0.35m) & STL-1001 \\
16/17 April 2010 &  Meade (0.35m) & STL-1001 \\
19/20 April 2010 &  Meade (0.35m) & STL-1001 \\
20/21 April 2010 &  Meade (0.35m) & STL-1001 \\

\hline
\end{tabular}
\end{center}
\end{table}

We also analysed ASAS-3 survey data (The All Sky Automated Survey; see Pojmanski 2002 for description of equipment and data pipeline) for BD+20~1790. Observations were carried out in the V-band during 2002-2009 (see Table~\ref{asaslog} for details). A Discrete Fourier Transformation was applied in order to estimate with better accuracy the known photometric period, and a dominant periodicity of $P_{rot}$ = $2.\!\!^{\rm d}79673 \pm 0.\!\!^{\rm
d}000075$ was revealed. Sample light curves, folded with this period are displayed on Fig.~\ref{fig:phase}.\\ 

A comparison of the eight-year light curves revealed that light curve minima occur at similar phases as shown in Fig.~\ref{fig:phase} and Fig.~\ref{fig:terskol}, indicating a configuration of photospheric activity that is very persistent in its location on the star. This suggests the presence of persistent dominant activity centered on a fixed longitude on the star. Stable active regions at the same longitude are also observed in RS CVn binaries (Howard 1996). The level of activity, size of the active region or number of spots in the active region is clearly variable with time, since the amplitude variability of the light curves is not constant. We find that $\Delta$V varied from 0.145 to 0.22 (see Fig.~\ref{fig:phase}).
The technique of Doppler imaging (Vogt 1983 \& 1987) enables starspots to be resolved on the surface of rapidly rotating stars, but requires sufficient resolution elements across the rotation profile to enable reliable reconstruction of spot distributions. As we noted in Hernan-Obispo et al. (2010), with a $v{\rm sin}i\sim$10 km s$^{-1}$, BD+20 1790 is not amenable to this technique. In addition, it is possible to obtain an estimate of the spot pattern using lightcurves alone. However, while simultaneous light curves at multiple passbands can in theory help to constrain the contrast of spots, the problem is highly degenerate with one photometric passband for single stars, even with image regularisation such as maximum entropy.

\subsection{Spectropolarimetric data}

\begin{figure}
   \centering
\includegraphics [scale=0.35]{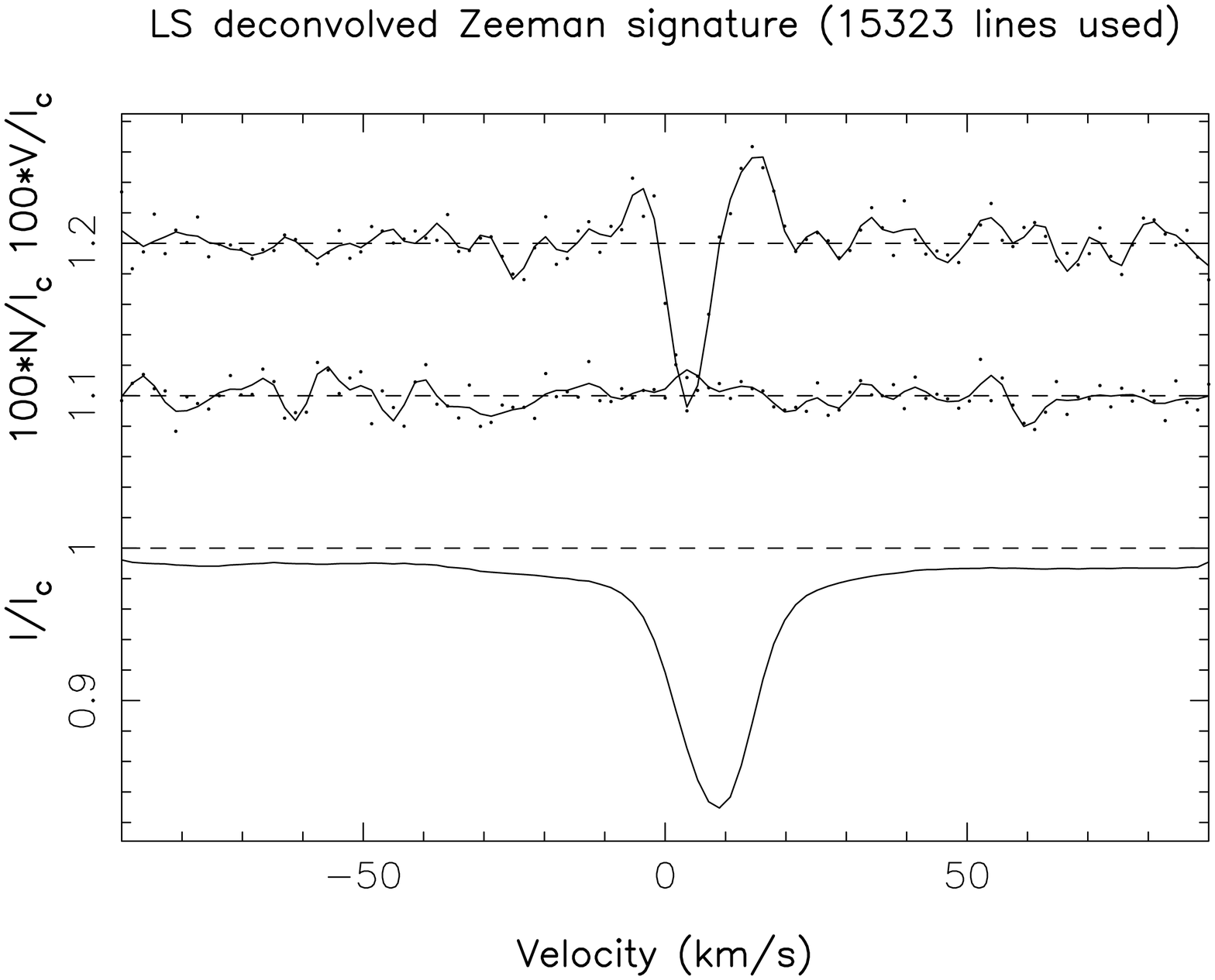}
\includegraphics [scale=0.35]{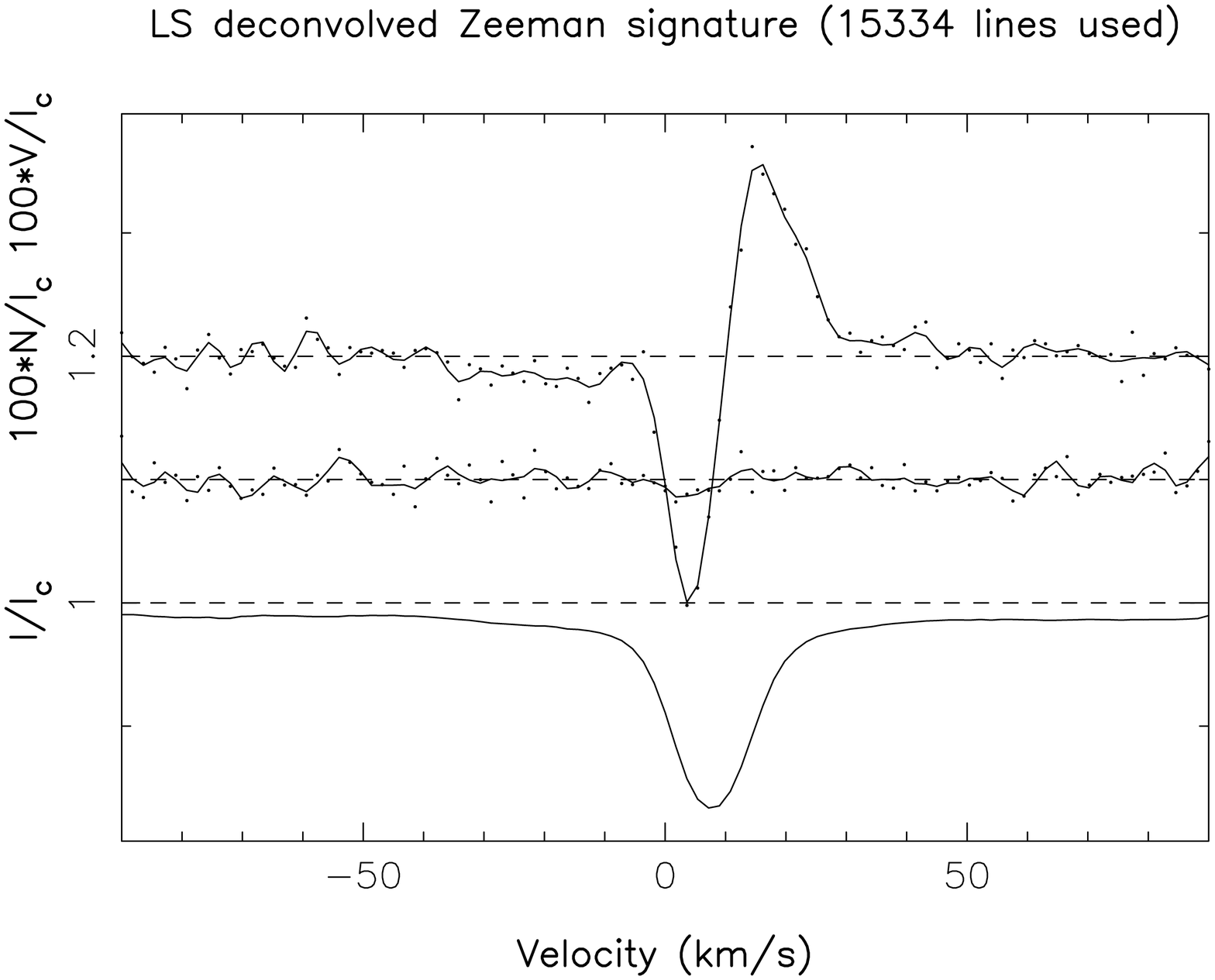}
       \caption{LSD profiles of BD+20 1790 for 2013 April 22 (top panel) and 2013 May 04 (bottom panel). For each panel, the upper plot shows the Stokes V LSD profile (normalized to the continuum level, expanded by 100 times and shifted up by 0.2), the middle plot shows the Null LSD profile (again normalized expanded by 100 times and shifted up by 0.1) and the lower plot shows the Stokes I normalized LSD profile.}
        \label{fig:narval}
\end{figure}

     BD+20 1790 was observed with the NARVAL spectropolarimeter (Auri\`ere 2003) at the Bernard Lyot Telescope of Pic du Midi Observatory. The instrument was used in its polarimetric configuration, and two spectra of the star were collected, on 2013 April 22 and May 04. Each observation provides us with both intensity (Stokes I) and circularly polarized (Stokes V) measurements, in a search for Zeeman signatures of surface magnetic fields. The data reduction and analysis was a strict duplicate of the one proposed by Marsden et al. (2013) for other cool active stars. The Least-Squares-Deconvolution cross-correlation technique (hereafter LSD, Donati et al. 1997) was applied to both Stokes I and V spectra, using a line-mask computed for stellar atmospheric parameters close to those of the star. By doing so, about 15,000 spectral features were simultaneously incorporated in the analysis, and both resulting LSD pseudo-line profiles display clear Stokes V signatures that we interpret as the spectral imprint of the stellar photospheric field through the Zeeman effect (see Fig.~\ref{fig:narval}). Applying the centre-of-gravity technique to estimate the line-of-sight projection of the stellar magnetic field (Rees \& Semel 1979), we derive values of $-19.6 \pm 2.3$~G and $-67.8 \pm 2.5$~G for April 22 and May 04, respectively. These field values are in global agreement with measurements reported by Marsden et al. (2014) for stars of similar $R^{\prime}_{HK}$ levels, and show that BD+20 1790 is an excellent possible target for magnetic mapping through Zeeman-Doppler Imaging methods.

\section{Results}

We analyse the RV data and various activity indices using samplings of posterior probability densities of the statistical models. We use the adaptive Metropolis algorithm (Haario et al. 2001) that is well suited for analyses of RV data because it provides representative samples from the posterior probability density in terms of close-Markovian chains regardless of the initial choice of the parameter vector or its covariance matrix (e.g. Tuomi 2011; Tuomi et al. 2011). To speed up the convergence of these chains, the initial parameter vectors in these samplings were set according to periodicities corresponding to the strongest peaks in the Lomb-Scargle periodograms (Lomb 1976; Scargle 1982) of the data.

When calculating the posterior probabilities of the different models, we use a sufficient burn-in period to allow the chain to converge to the posterior density. After this burn-in period, we fix the proposed density of the sampling, making the process the Markovian Metropolis-Hastings algorithm (Metropolis et al. 1953; Hastings 1970), and use the method of Chib \& Jeliazkov (2001) to calculate estimates for the marginal likelihoods and the corresponding model probabilities.

We model the RVs with a statistical model containing periodic signals and reference velocities for each telescope-instrument combination. We assume that the measurements have a Gaussian distribution with an unknown variance. As prior densities of the parameters of our statistical models, we use the same functional forms as in Ford \& Gregory (2007).

\subsection {Bayesian analysis of RV data}

According to our results, the FIES, FOCES, HERMES, and SARG radial velocities data are consistent with one another. They all contain a strong periodicity at 7.78 days corresponding to the proposed planetary signal reported in Paper I {\bf (see details at Sect. 4. 2.)}. However, as noted by F10, this periodicity is not present in the CORALIE radial velocities despite the fact that its large amplitude of 872 ms$^{-1}$ should have enabled a detection. We analysed the CORALIE-RV data by performing Markov Chain Monte Carlo (MCMC) samplings of the parameter space of the one-Keplerian model. We especially searched the period-space between 1 and 10 days by limiting the parameter space of the period to this interval. However, we could not find clear probability maxima corresponding to periodic signals. Instead, we found the period-space to consist of several local maxima out of which none was found clear enough to constrain the corresponding periods. We also analysed the CORALIE RV-data together with the combined FIES, FOCES, HERMES and SARG data by performing MCMC samplings of the parameter space. While the strongest signal with a periodicity of 7.78 days was still found in this data set composing of RV observations, the corresponding one-Keplerian model resulted in a poorer fit to the CORALIE data than a model without any signals. This was especially indicated by studying the periodogram of the CORALIE data and their residuals - the residuals of the one-Keplerian model showed a strong power at 7.78 days while there were no significant powers in the original measurements. 

Therefore, the CORALIE RV-data were clearly inconsistent with the rest of the measurements.

\subsection {Searching for activity-related periodicities in RV data}

\begin{figure}

\includegraphics[scale=0.38,clip]{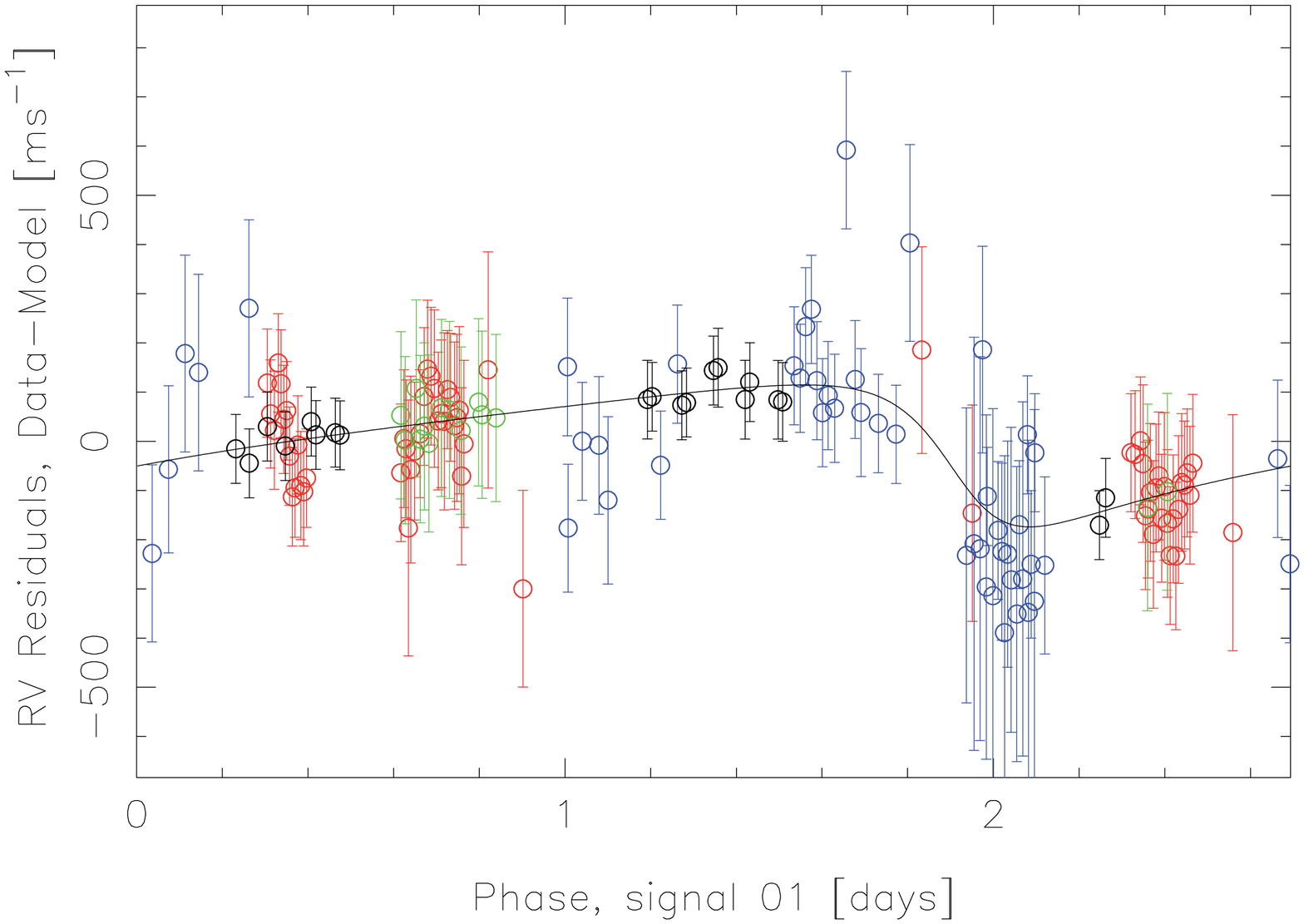}

\includegraphics[scale=0.38]{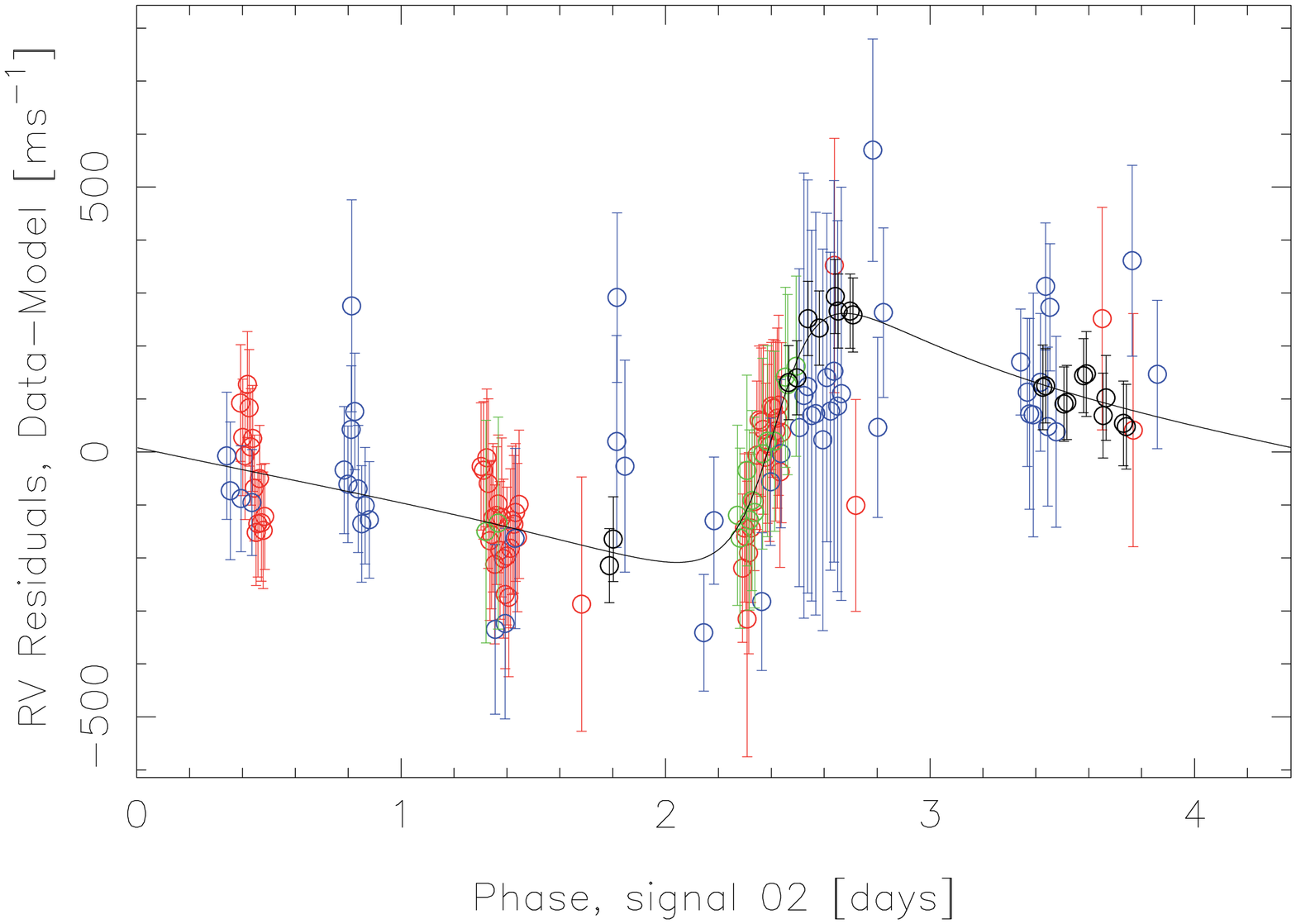}

\includegraphics[scale=0.38]{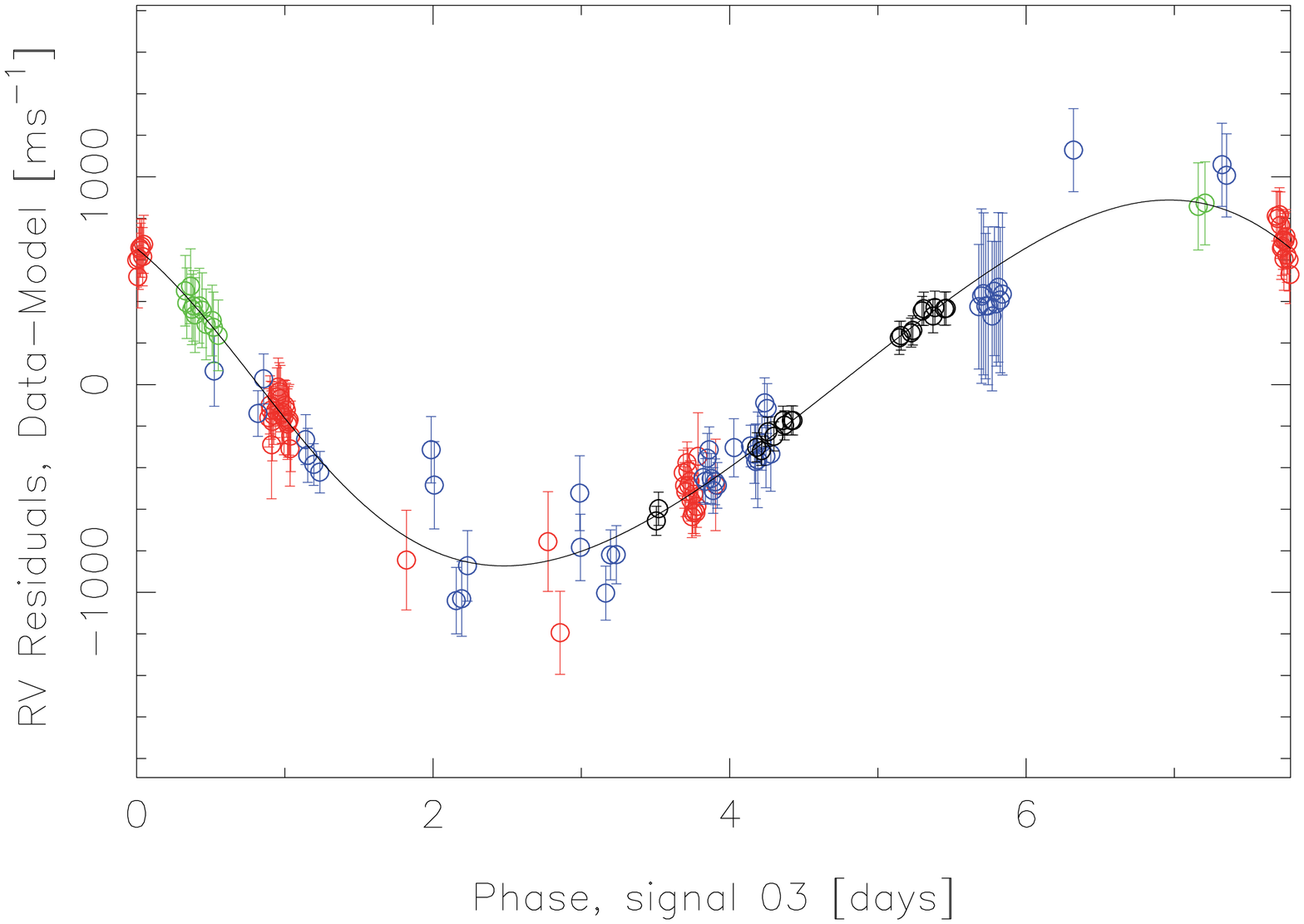}
\caption{The three signals in the combined FIES (green), FOCES (blue), SARG (red), and HERMES (black) radial velocities corresponding to the 2.69 (top), 4.36 (middle), and 7.78 (bottom) periodicities with the other signals subtracted.}
  \label{fig:mikkofit}
\end{figure}

Because of the active nature of BD+20~1790, we also searched for activity-related periodicities in the RV. While additional periodicities were not statistically significant in the individual FIES, FOCES, HERMES, and SARG radial velocities with threshold probability of 99\%, the combined data set was found to have two additional periodicities at 2.69 and 4.36 days, corresponding to RV amplitudes of 156 and 165 m/s. We calculated the Bayesian model probabilities for models with 0-3 periodic signals. These probabilities are $4.3 \times 10^{-97}$, $3.8 \times 10^{-27}$,  $4.1 \times 10^{-9}$ and $\sim 1.0$, respectively, for the combined data. According to these probabilities, the periodicity at 7.78 days is very clearly present in the data, the 4.36 day period is the second strongest - the posterior probability of the two-periodicity model is roughly $1.0 \times 10^{18}$ times more probable than that with a single periodicity. The three-periodicity model receives the greatest posterior probability of $1$\,$-$\,$4.9 \times 10^{-9}$, which means that we can detect the 2.69 day periodicity in the RVs data with strong confidence, likely corresponding to the stellar rotation period. These signals are shown in Fig.~\ref{fig:mikkofit} for the combined data set. 

The 4.36 day signal corresponds with the synodic period of the star-planet system ($P_{\rm syn}^{-1} = | P_{\rm rot}^{-1} - P_{\rm orb}^{-1} | $). Because the rotation of the planet is not synchronized with the stellar rotation, it allows us to separate the modulations from signals at different periods (orbital, rotational, synodical). The synodic period is the time between two subsequent crossings of a given meridian to the sub-planetary point, i. e., between two successive alignments of an active surface region and the planet.

The stellar jitter is not expected to be sinusoidal, because we are comparing data from different epochs with different levels of activity, and also the effect of flares introduces a stochastic noise into the data, that could swallow a sinusoidal fit.

The Bayesian analysis of the CORALIE-RV data did not reveal the 7.78 day signal or the one claimed by F10 close the photometric period.

\subsection {Bayesian analysis of chromospheric activity indices}

The aim to perform a Bayesian analysis of the activity indicators, apart from finding periodicities, was to check if the 7.78 day signal is a real period. If the 7.78 day signal is a misleading signal, an artefact due to the manner in which the observations were made, it should be detected in every parameter that we measured from the spectra. To test this, we carried out a simultaneous analysis of the activity indices, measured from the same spectra as the RV. Two analyses were performed, first considering the whole data sets without CORALIE, and a second for CORALIE data only. 

 \begin{figure*}
   \centering
   \includegraphics[scale=.37]{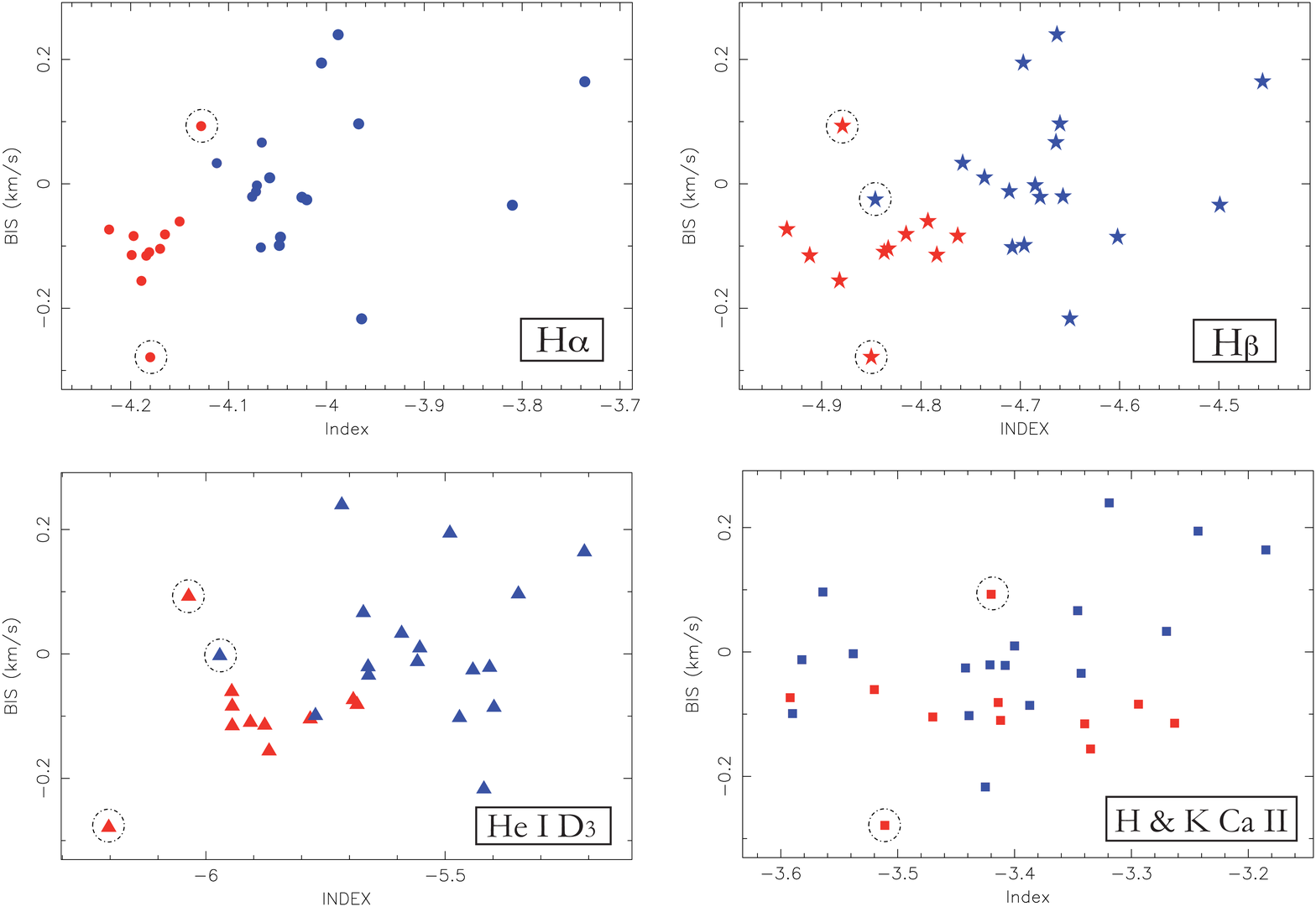}
  \caption{Activity index vs. BIS for CORALIE data. Blue symbols represent flare affected indices. Red symbols represent the data without flare activity. Circles are for H$\alpha$, stars for H$\beta$, triangles for He I $D_{3}$, squares for Ca~{\sc ii} H \& K. It can be seen that the scatter for the BIS is higher when a flare event occurs. Points that are marked with a dashed circle present a peculiar behaviour that is discussed in the text. Error bars for indices are about 0.001 and for BIS about 0.0001 km/s.}
        \label{fig:indices}
   \end{figure*}

In the first case, the Bayesian analysis of the H$\alpha$ index revealed a strong modulation with 4.36 days and moderately strong one with 2.7 days. For H$\beta$ we found a strong signal at 4.30 days, likely corresponding to the 4.36 day signal, and another strong signal at 2.62 days. As expected the 7.78 day period is only seen in the RV data. It has no photometric and no chromospheric counterpart. The variations of the Ca~{\sc ii} IRT indices showed similar trends. We cannot find significant signals at He I $D_{3}$. We performed the Bayesian analysis for the non-flare affected data of H$\alpha$ and He I $D_{3}$ indices. We found the 2.7 and 4.36 day signal in the H$\alpha$ index but not in the He I $D_{3}$. 

We studied mainly the variability of the H$\alpha$, H$\beta$ and Ca~{\sc ii} IRT indices, instead of the traditional Ca~{\sc ii} H \& K, because of their low S/N for mid-K stars that are redder and fainter at these wavelengths. The mean S/N in our runs for every indicator are: Ca~{\sc ii} H \& K $\sim$ 9, H$\alpha$ $\sim$ 56, H$\beta$ $\sim$ 36 and for Ca~{\sc ii} IRT $\sim$ 60.

The Bayesian analysis of the CORALIE activity indices showed trends of some periodicities, but we cannot quantify them because their amplitudes do not exceed the average level of random variations in the data.

\subsection {Jitter from flares in the CORALIE data}

A high rate of flare events is detected in the CORALIE data: 16 of a total of 26 spectra show flare activity. Table~\ref{flares} summarizes the number of flares found in all the runs, and in the case of CORALIE data, indicating the number of flares present in both data sets. 

The relationship between the bisectors of the cross-correlation function (CCF), and RV is widely used to discriminate if the variations of RV are due to stellar activity or a planetary companion. To quantify the changes in the shape of the CCF, the bisector inverse slope (BIS; see Queloz et al. 2001 for details about the method) is computed.

As we reported for first time in Paper I, BIS are strongly sensitive to flares (for further details see Sect.~4. 3 and Fig.~12 at Paper I). Set 2 is dominated by flare activity (6 of 8), whereas more than half of the measurements of Set 1 present a flare event. Using the same analysis as we carried out in Paper I, we investigated whether the CORALIE BIS data could be contaminated by flares. We show the plots of BIS vs. different activity indices in Fig.~\ref{fig:indices}, where flare-affected and non-affected are pointed out with different colors (red for quiescent state and blue for flare state). The scatter in the case of flare-affected BIS is clearly appreciable in the case of the Balmer lines, and He I $D_{3}$. 
 
\begin{table}
\begin{center}
\caption{\label{flares} Number of flares for the CORALIE run (both Set 1 and Set 2), HARPS-N all runs, and for the rest of the runs (FIES, FOCES, MERCATOR and SARG runs, see Table~\ref{runs} on this paper and Table~2 at Paper I for further details).}
\begin{tabular}{lccc}
\hline \hline Run & $Flare$  & $Total$ & $ Flare -Rate$ \\

\hline
CORALIE Set1 & 10 & 18 & 55\% \\
CORALIE Set2 & 6 & 8 & 75\% \\
\hline
CORALIE All & 16 & 26 & 61\% \\

HARPS-N & 4 & 8 & 50\% \\
Rest of runs & 16 & 27 & 59\% \\
\hline
Total & 36 & 61 & 59\% \\

\hline \hline
\end{tabular}
\end{center}
\end{table}

\begin{figure}[h!]
   \centering
   \includegraphics[scale=.34]{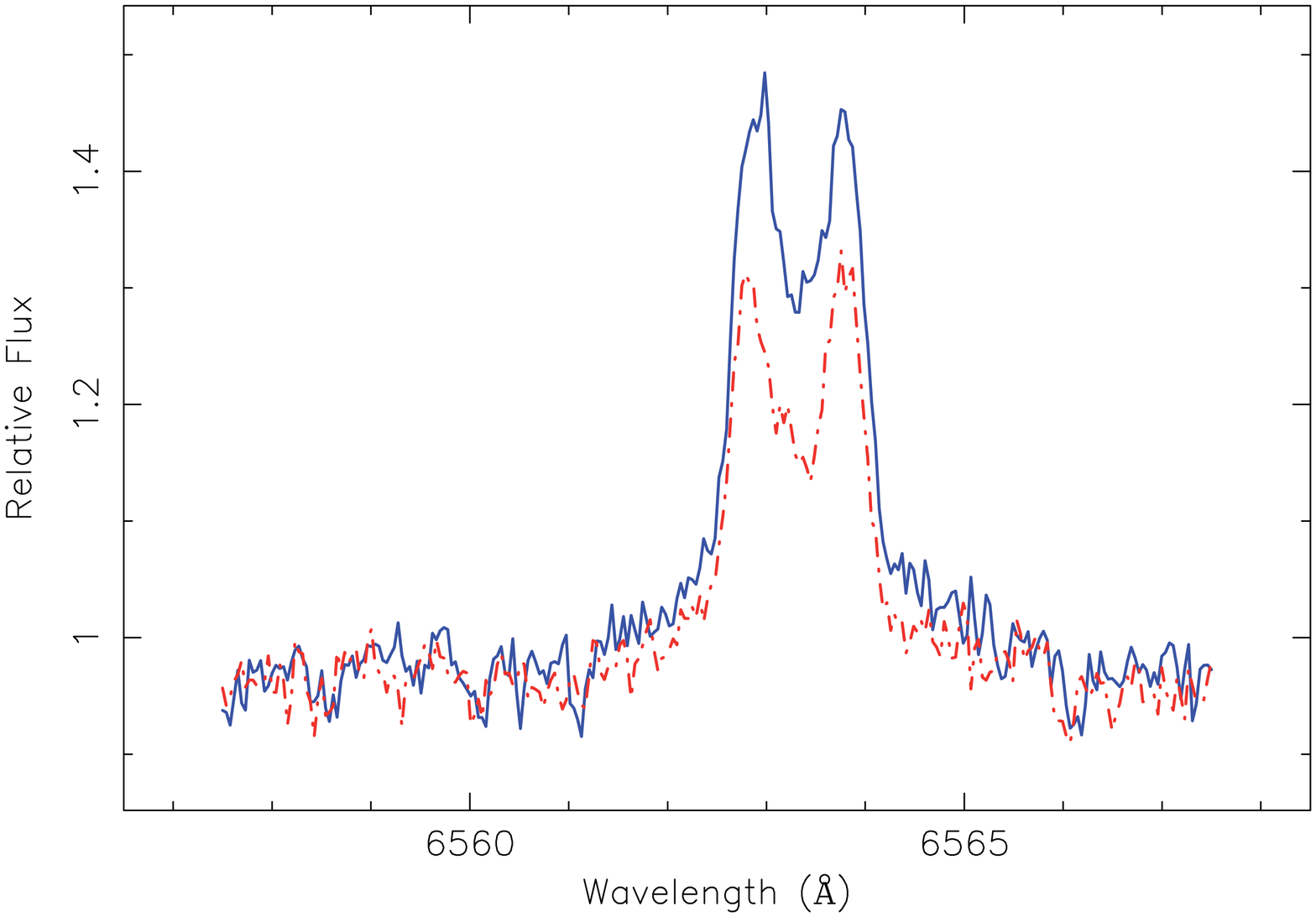}
   \includegraphics[scale=.34]{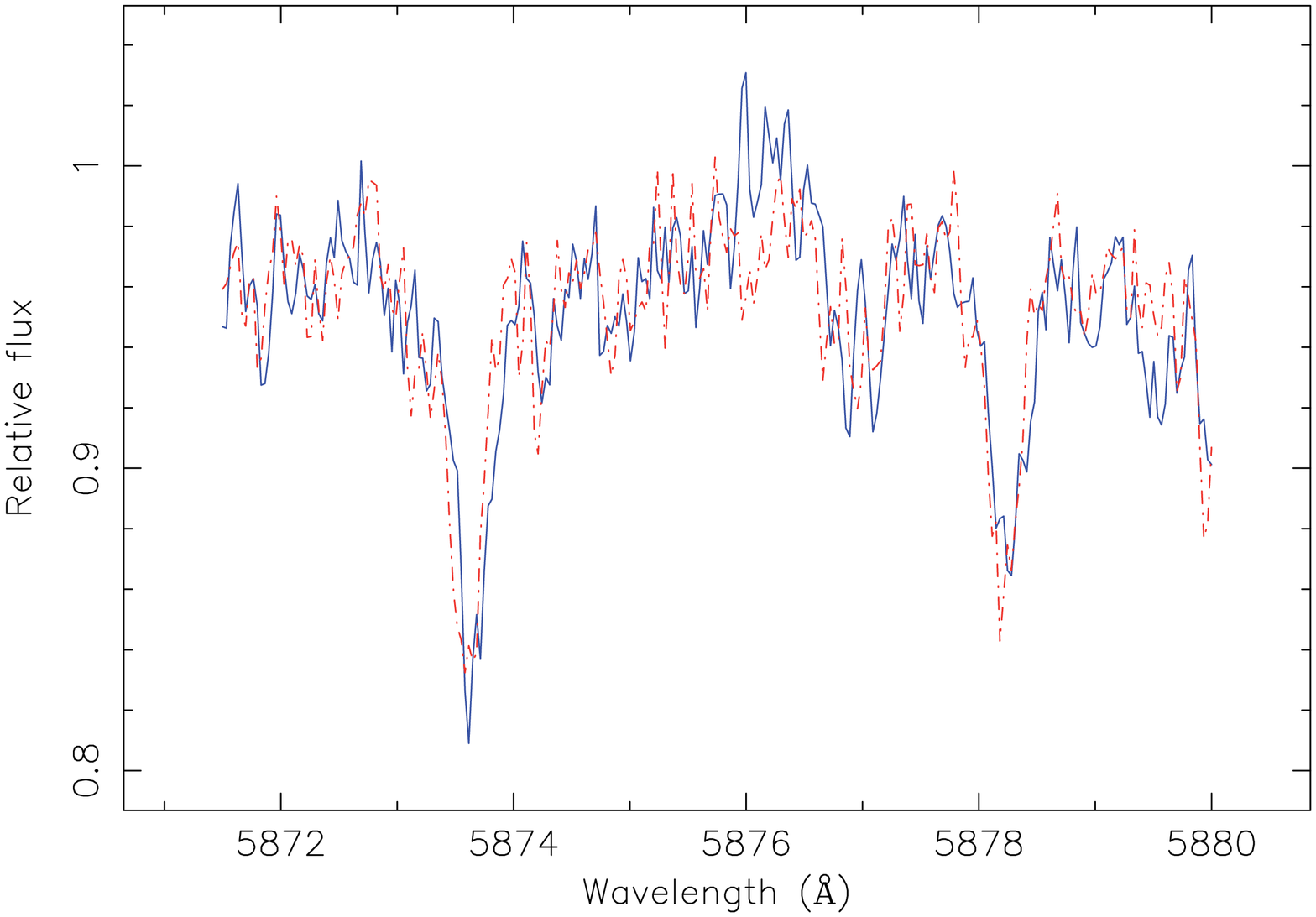}
        \caption{\textbf {Top}: H$\alpha$ line for the blue point marked in Fig.~\ref{fig:indices}. \textbf{Bottom}: He I $D_{3}$ line for the blue point marked in Fig.~\ref{fig:indices}. The red dashed line is indicating the quiescent state.}
        \label{fig:puntoazul}
   \end{figure}

The three points marked with a dashed circle present a peculiar behaviour. The two quiescent-state points show unusual low activity level in comparison with the rest. The blue point marked with a dashed circle shows an apparently similar index to the value for the nearest quiescent point but appears clearly to be a in flare state in Fig.~\ref{fig:puntoazul}. If we plot the H$\alpha$ and He I $D_{3}$ lines for this point vs. the quiescent it can be seen that the star is flaring (See Fig.~\ref{fig:puntoazul}). 

Fig.~\ref{fig:indices} also illustrates the response in height of the chromosphere when a flare occurs, as the different activity indices form at different heights (and therefore at different temperature). He I $D_{3}$ forms in the upper chromosphere, Balmer lines in the middle chromosphere and Ca~{\sc ii} H \& K lines in the middle and low chromosphere. Apart from this, every emission line decays with a different time scale, with the Ca~{\sc ii} H \& K lines being the more delayed. Ca~{\sc ii} H \& K have a slower response to the flare than the He I $D_{3}$ line, that rises rapidly and the Balmer lines, that follow He I $D_{3}$.

\begin{figure}[h!]
   \centering
  
   \includegraphics[scale=.34]{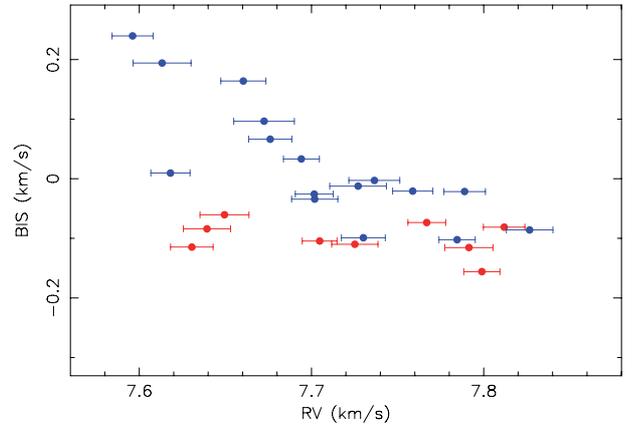}
	\caption{RV vs. BIS for CORALIE data. Blue symbols represent flare-affected data for CORALIE both runs. Red symbols represent the data without flare activity. Error bars for BIS are about 0.0001 km/s.}
        \label{fig:biscoralie}
   \end{figure}
   
Considering all this information, we plotted in Fig.~\ref{fig:biscoralie} the RV vs. BIS for the CORALIE data, marking flare (blue symbols) and no-flare (red symbols) states. This figure corresponds to Fig.~4 of F10 paper. As can be seen in Fig.~\ref{fig:biscoralie}, the RV data contaminated by flares anti-correlate with BIS (Pearson correlation coefficient r= -0.808229), whereas the data without flares show a lack of correlation. This indicates that, as presented above, the BIS are strongly sensitive to flares, and also pointed out that the BIS are not a good discriminator when the data are flare-contaminated. Also we should analyse with care these RV measurements affected by flares.\\

\subsection{Fake star model and signals detected at CORALIE data}

In order to test if we are able to recover the planetary signal with software similar to the CORALIE pipeline, we generated a fake star model, consisting of spot jitter and a Keplerian signal. The aim of this experiment was to try to recover the signals that we introduced in the fake data.

While image reconstruction from light curves can yield an indication of the dominant spot longitude, latitudinal information is unreliable for the reasons outlined in Sect 3.2. However, since images can nevertheless be derived, they can then be used in a forward modelling sense to generate spectroscopic line profiles. This procedure is detailed in Barnes et al. (2011) for M dwarfs, where we used synthetic spot model images rather than data-derived images. By taking the image solution for the LT 2007 and 2008 light curves, we generated spectral lines that are distorted by starspots for the times at which BD+201790 was observed spectroscopically (i.e. 145 observations). A rotation period of 2.8 days was used. The line profiles were then convolved with a synthetic spectrum for a 4250 K star, covering the spectral range of the observations. Finally, the proposed planetary signature was injected with a 7.78 day period. 

\begin{figure}[h!]

\includegraphics[scale=0.38]{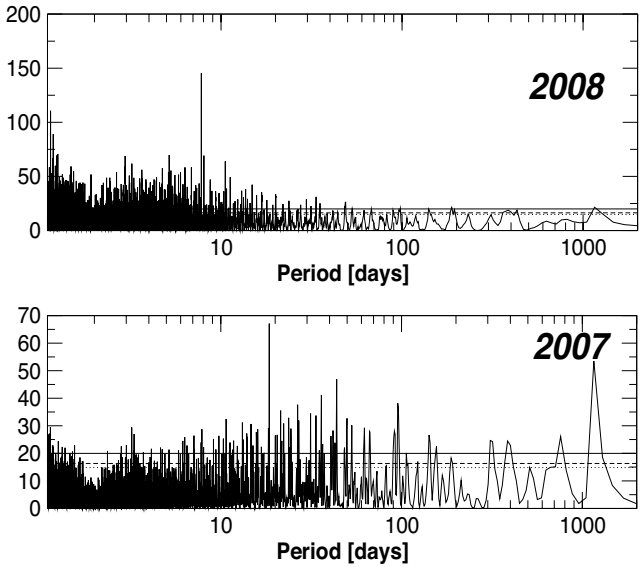}
	\caption{Periodograms for LT 2007 (down) and LT 2008 (up) for the fake star model.}
     \label{fig:fakestar}
   \end{figure}

A period search was carried out in order to recover the periodic signatures, using HARPS-TERRA software (Anglada-Escud\'e \& Butler 2012). We found that for the first LT epoch (LT 2007), where the level of stellar activity was high, we could not recover any of the signals, neither the orbital (7.78 days) or photometric (2.8 days) periods, as can be seen in the periodogram in Fig. ~\ref{fig:fakestar}. These results are thus in agreement with our Bayesian analysis of CORALIE data where we did not recover any signal, but instead we found a high level of stellar activity. 

In the second epoch, LT 2008, where the star presented a lower level of activity, we recovered the 7.78 day signal, but not the 2.8 day signal. It is possible that the planet signal is masked by the activity of the star when the activity level is sufficiently high, while it can be recovered when activity levels are lower. Our experiment of the fake star data attests that actual pipelines could yield spurious results when data from very active stars is analysed without careful consideration of activity-affected measurements. Since most of the exoplanet RV searched have avoided active potential host stars, the RV pipelines developed have not taken into account the problems that can arise from activity. Our results do not invalidate any planetary detection claimed using the CORALIE or other software pipelines. Only shows that, when applying software pipelines to precision RVs affected by high levels of stellar activity, all the signals present in the data may not necessarily be recovered, even signals related to the stellar features.  

In a forthcoming paper we will explore this in detail, Hern\'an-Obispo et al. (in prep.).

\subsection {HARPS-N vs. CORALIE data}  
    
\begin{figure}
   \centering
     \includegraphics[scale=.38]{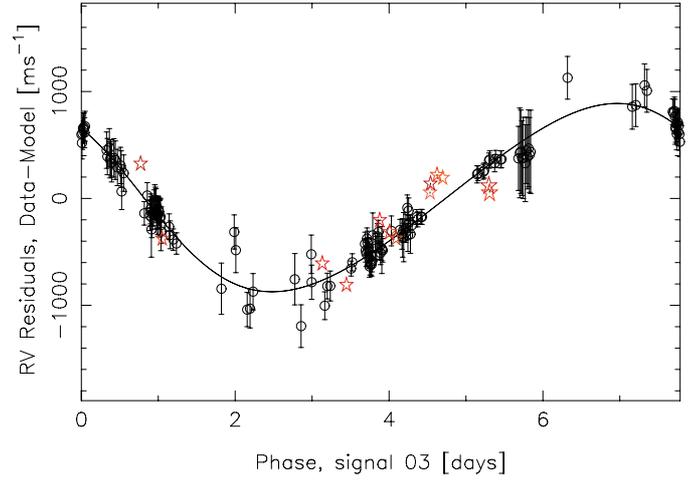}
	\caption{Radial velocity of all runs and HARPS-N. Red stars represent HARPS data. Black symbols represent the rest of runs. Error bars for HARPS data are about 0.004 km/s.}
        \label{fig:fit_orbit}
   \end{figure}

We observed with HARPS-N data to further monitor RV variability of BD+20 1790 and compare with CORALIE data. Since CORALIE was the precursor spectrograph to HARPS-N, and both have similar quality, the comparison is justified. Unfortunately we cannot provide an orbital solution using only the HARPS-N data, because there are insufficient observations that do not cover the full rotational period. Due to HARPS-N errors being about two orders of magnitude lower than the rest of our RV data, the Bayesian analysis presented problems.

Nevertheless, we can plot the HARPS-N data over the RV curve obtained with previous runs, calculating the phase of each point. As can be seen in Fig.~\ref{fig:fit_orbit}, HARPS-N data agree with the curve solution.\\

\begin{figure*}
   \centering
  \includegraphics[scale=.37]{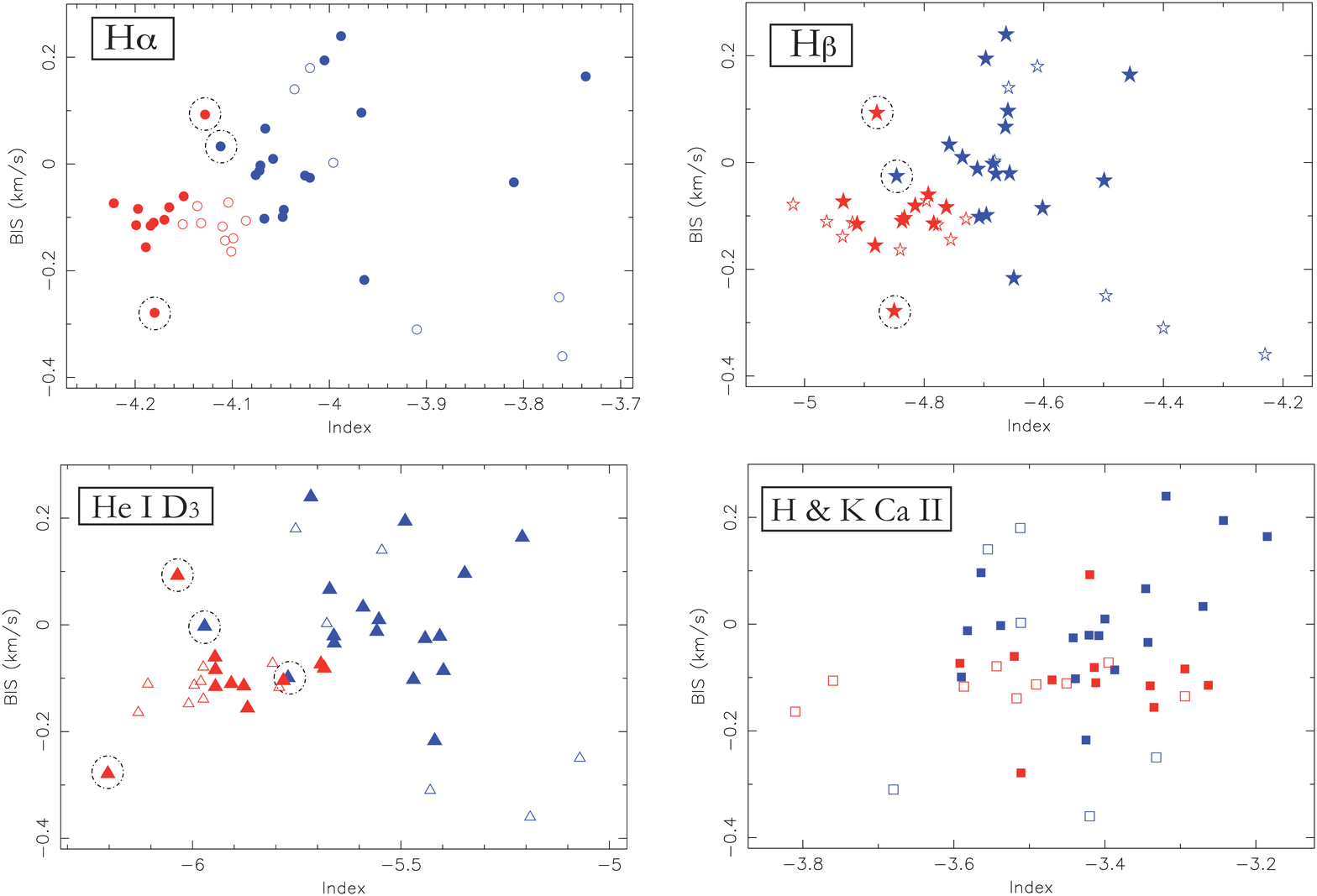}
  \caption{Activity index vs. BIS for CORALIE and HARPS-N data. Circles are for H$\alpha$, stars for H$\beta$, triangles for He I $D_{3}$, squares for Ca~{\sc ii} H \& K. Blue symbols represent flare affected indices. Red symbols represent the data without flare activity. Filled circle are for CORALIE data. The rest for HARPS-N data. It can be seen that the scatter for the BIS is higher when a flare event occurs. Points that are marked with a dashed circle present a peculiar behaviour that is discussed in the text. Error bars for indices are about 0.001 and for BIS about 0.0001 km/s.}
        \label{fig:indicesharps}
   \end{figure*}

We determined the HARPS-N activity indices in the same way as for the rest of runs. Four flare events were detected, two of them extremely large, as was shown in Fig.~\ref{fig:megaflare}. 

Fig.~\ref{fig:indicesharps} compares the activity indices vs. BIS for HARPS-N (open symbols) and CORALIE (filled symbols) data. It is clearly seen that HARPS-N and CORALIE activity data are consistent. As shown for CORALIE data, we can see an appreciable scatter in the case of flare-affected BIS.

\begin{figure}
   \centering
  
   \includegraphics[scale=.34]{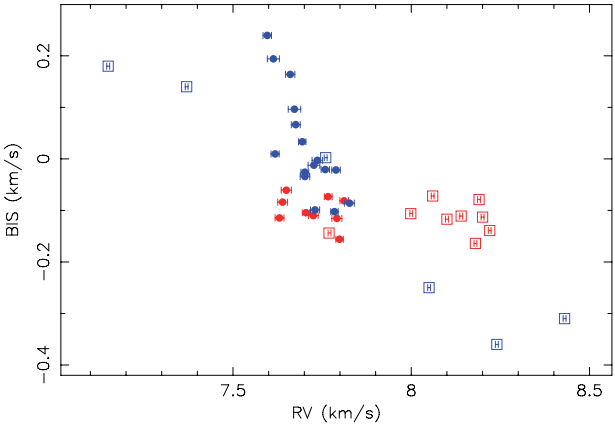}
	\caption{RV vs. BIS for CORALIE  and HARPS-N data. Blue symbols represent flare-affected data for CORALIE both runs. Red symbols represent the data without flare activity. Filled symbols are for CORALIE data and open symbols are for HARPS-N. Error bars for BIS are about 0.0001 km/s.}
        \label{fig:bisharps}
   \end{figure}

In Fig.~\ref{fig:bisharps} HARPS-N BIS vs. RV data superposed to CORALIE data is shown. Again, anti-correlation of the flare-affected data (plotted in open blue circle in the figure) is clearly seen, whereas the non-flare-affected data show a lack of correlation (red symbols). This figure also shows that the RVs peak-to-peak amplitude for HARPS-N is about 1300 m s$^{-1}$. The HARPS-N data collected span approximately one half of the orbital period. In the case of CORALIE data, the RVs peak-to-peak amplitude is more than three times smaller than HARPS-N, although CORALIE data cover several times the orbital period.

In light of RV Bayesian analysis and Figs.~\ref{fig:fit_orbit}, \ref{fig:indicesharps} and \ref{fig:bisharps}, we can conclude that the CORALIE-RV data are inconsistent with all the other data sets, including results obtained by similar instrument, HARPS-N.

\section{New orbital solution}

With all these considerations, we computed again the orbital solution for the RV data, with the constraints of a second and a third periodicity signal (actually the photometric and synodic activity signals). Orbital elements are given at Table~6. We plotted the results of our posterior samplings in terms of maximum a posteriori (MAP) estimates and the corresponding 99\% Bayesian credibility sets (BCS) (see e.g. Tuomi \& Kotiranta 2009) in Table~6. The parameters $\gamma_1$, ...,$\gamma_4$ denote the reference velocities of FIES, FOCES, HERMES, and SARG radial velocities. The parameters $\sigma_{I,1}$, ..., $\sigma_{I,4}$ are similarly to the noise parameters that were used to weight the uncertainty estimates of the four data sets by multiplying each uncertainty estimate with a corresponding parameter. The resulting coefficients $\sigma_{i,l}=\sigma_{i} *\sigma_{l}$, where $\sigma_{i}$ is the uncertainty estimate of the $i$th measurement, were then treated as standard deviations of the measurement $i$ made using instrument $l$ by assuming that the measurements had Gaussian densities. A value of unity for these noise parameters means that the uncertainty estimates of the corresponding data set were estimated correctly, whereas values below (above) unity mean that the uncertainties were overestimated (underestimated).
We remark that the values of Table~6 should be revised when a follow up allows us to coverage of the complete orbit.

\begin{table*}
\center
\label{planet2}
\caption{The MAP estimates and the 99\% BCS's of the one- and three-periodicity models of the combined FIES, FOCES,  HERMES, and SARG radial velocities. First block shows orbital elements of BD+20 1790 b}
\begin{tabular}{lcc}
\hline \hline
Parameter & 1 Keplerian & 3 Keplerian  \\
& MAP BCS & MAP BCS \\
\hline
$P$ [days] & 7.78287 [7.78211, 7.78364] & 7.78429 [7.78367, 7.76508]\\
$e$ & 0.22 [0.13, 0.29] & 0.13 [0.06, 0.20] \\
$K$ [ms$^{-1}$] & 905 [811, 998] & 872 [816, 929]  \\
$\omega$ [rad] & 4.58 [4.09, 4.96] & 1.61 [0.87, 2.29] \\
$t_{0}$ [rad] & 3.97 [3.26, 4.68]  & 0.73 [0.00, 1.47] \\
$M_2 \sin i$ [M$_{\rm Jup}$] & 6.37 [5.02, 7.72] & 6.24 [4.9, 7.58] \\
$a$ [AU] & 0.066 [0.060, 0.071] & 0.066 [0.060, 0.071]  \\
\hline
$P_{1}$  [days] & -- & 2.69367 [2.69327, 2.69395] \\
$K_{1}$ [ms$^{-1}$] & -- & 117  [63, 170] \\
$P_{2}$  [days] & -- & 4.36477 [4.36457, 4.36492] \\
$K_{2}$ [ms$^{-1}$] & -- & 187 [135, 258] \\
\hline
$\gamma_{1}$ [ms$^{-1}$] (SARG) & 157 [81, 219] & 64 [-9, 137] \\
$\gamma_{2}$ [ms$^{-1}$] (FOCES) & 164 [62, 246] & 138 [73, 212] \\
$\gamma_{3}$ [ms$^{-1}$] (FIES) & -21 [-182, 139] & -414 [-503, -290] \\
$\gamma_{4}$ [ms$^{-1}$] (HERMES) & -521 [-670, -350] & -165 [-270, -27]\\
$\sigma_{I,1}$ [ms$^{-1}$] (SARG) & 0.60 [0.36, 1.05] & 0.42 [0.23, 0.65]\\
$\sigma_{I,2}$ [ms$^{-1}$] (FOCES) & 1.74 [0.92, 3.06] & 0.84 [0.44, 1.53] \\
$\sigma_{I,3}$ [ms$^{-1}$] (FIES) & 0.43 [0.11, 1.32] & 0.033 [0.009, 0.090] \\
$\sigma_{I,4}$ [ms$^{-1}$] (HERMES) & 0.29 [0.11, 1.18] & 0.13 [0.05, 0.33] \\
\hline \hline
\end{tabular}
\end{table*}

\section {On possible star-planet-interaction}

\subsection {Star-planet interaction stage}
Through its orbit around the star, a close-in giant planet is embedded in the surrounding magnetic field of the star. One might expect not only the influence of the star over the planet (evaporation and strong irradiation, synchronization and circularization of the orbit, etc.) but also some impact on the star from the planet presence, in analogy to what is observed in binary stars. It is well established that binary stars are more active than the same spectral type and age single active stars (Glebocki et al. 1986, Ayres \& Linsky 1980). Extensive literature has shown significant chromospheric and coronal activity (Siarkowski et al. 1996, Ferreira 1998), and spectacular flare activity, in some cases occurring even in the inter-binary space and during periastron (Graffagnino et al. 1995, Massi et al. 2002, Massi et al. 2008, Salter et al. 2010). Rotationally modulated flares have been found to occur in several binary systems (Garc\'ia-\'Alvarez  et al. 2003, Doyle et al. 1990).

\begin{figure}
   \centering
   \includegraphics[scale=.32]{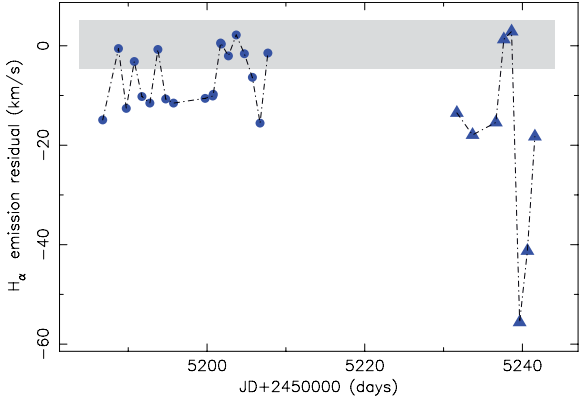}
  \caption{Residual emission for H$\alpha$ line of CORALIE runs vs. JD. Circles are for Set1 and Triangles for Set2. The gray box indicates quiescent state.}
        \label{fig:residuos}
   \end{figure}

\begin{figure}
   \centering
   \includegraphics[scale=.32]{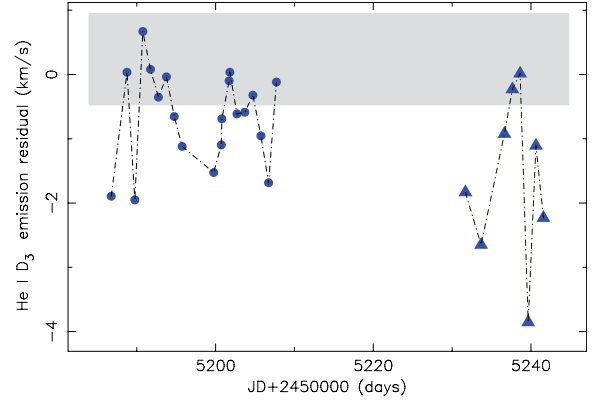}
	\caption{Residual emission for He I $D_{3}$ line of CORALIE runs vs. JD. Circles are for Set1 and Triangles for Set2. The gray box indicates quiescent state.}
        \label{fig:residuoshelio}
   \end{figure}

The possible effects that a close-in giant planet may have on its host star, called generically star-planet-interactions (hereafter SPI) are currently a widely debated field that gives rise to many open questions. As proposed by Cuntz et al. (2000), this SPI could have both a gravitational and a magnetic origin.
Observational efforts to identify and characterize these SPI signatures have been carried out including wavelength exploration ranging from radio (Zarka et al. 2007, Griessmeier et al. 2007, Hallinan et al. 2013) to X-Rays (Poppenhaeger et al. 2010,  2011, Kashyap et al. 2008, Pilliteri et al. 2011), through UV (Shkolnik  2013) to the optical domain, but with different and some times incompatible results. Shkolnik et al. (2003) first reported evidence of chromospheric activity modulated with the orbital period rather than the photometric for HD 179949 (see also Shkolnik et al. 2005, 2008). It is also suggested by Shkolnik et al. (2008) that SPI could be cyclic and oscillate between "on" and "off" states. Possible photospheric SPI have been suggested by Walker et al. (2008), Pagano et al. (2009), Lanza (2009, 2011), based on stellar fluxes that modulated with the orbital period or the synodical period of the star and planet system. In the X-Ray regime, the statistical studies of a possible correlation between the X-ray emission of stars vs. their planets parameters (like semi-major axis and mass), have revealed controversial results. Kashyap et al. (2008) suggested that stars with close-in planets are more active than stars with distant planets. On the contrary, Poppenhaeger et al. (2010) and Poppenhaeger et al. (2011) did not find any correlation between stellar activity and the mass or the semi-major axis. Theoretically, Cohen et al. (2009)  demonstrated the relation by magnetohydrodynamic simulations, showing that SPI may increase the X-ray luminosity. Canto Martins et al. (2011) and Krej\v cov\'a  \& Budaj (2012) extended to the chromosphere these statistical studies of a possible dependence of the stellar activity with both the semi-major axis and mass. Whereas Canto Martins et al. (2011) did not detect a difference in the chromospheric emission of late-type stars with and without planets, Krej\v cov\'a \& Budaj (2012) found statistically significant evidence that chromospheric emission increases with both the mass and semi-major axis of the planet. This suggests that close-in massive planets may affect the level of chromospheric activity of its host star.
Some authors suggested some excess of flaring activity in phase with the planet rotation (Shkolnik et al. 2008, Pillitteri et al. 2011, 2014).

A recent review by Lanza (2014) recaps the open questions in this field and the theoretical models made to explain the different observational results obtained.

\subsection{The BD+20 1790 case}
The results of the Bayesian analysis of the activity indices revealed that the activity is modulated strongly with the synodic period rather than the photometric as expected. Fares et al. (2010) suggested that the enhancement due to magnetospheric SPI is more likely to be modulated with the synodic period of the star+planet system. This leads us to suspect possible SPI in BD+20 1790.

Except for CORALIE data, the time sampling is poor, the rest of the data series cover only a fraction of the stellar rotation or planet orbital period. However the CORALIE data have an optimal coverage of both periods, enough to perform a detailed study of possible modulation of the activity indices, not only to discern suspicious trends from Bayesian analysis. We computed the residual emission for H$\alpha$ and He I $D_{3}$ lines, following Fares et al. (2010). Fig.~\ref{fig:residuos} and Fig.~\ref{fig:residuoshelio} show the variation vs. time (Julian Date, JD) for both indices. Circle and triangle symbols represent the first and second set respectively. By simple visual inspection of the plots, data affected by flares are clearly distinguished from the ones without flares (quiescent state is marked with a gray area).

Due to the high occurrence of flare events it is difficult to disentangle any periodicities in the system. When removing the flare-affected data the number of points is not enough for assessing any discrimination. Notwithstanding, we cannot exclude the possibility that the activity was modulated with the synodic period nor the photometric one, since flaring prevents effective disentangling periodicities.
On the other hand we cannot exclude the hypothesis that these flare events could be related to SPI.

\begin{figure}
  \centering
  \includegraphics[scale=0.5] {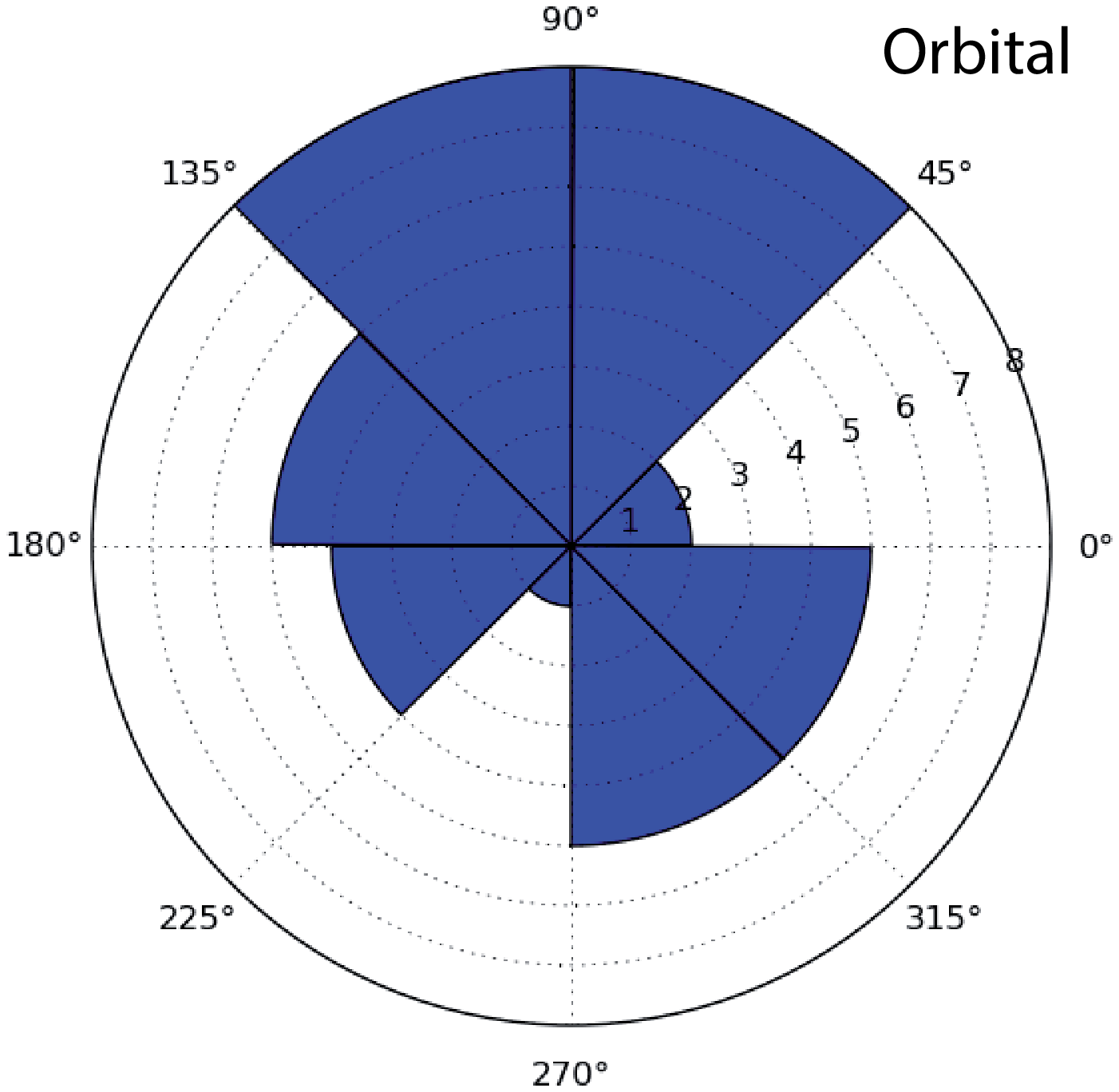}
   \includegraphics[scale=0.5]{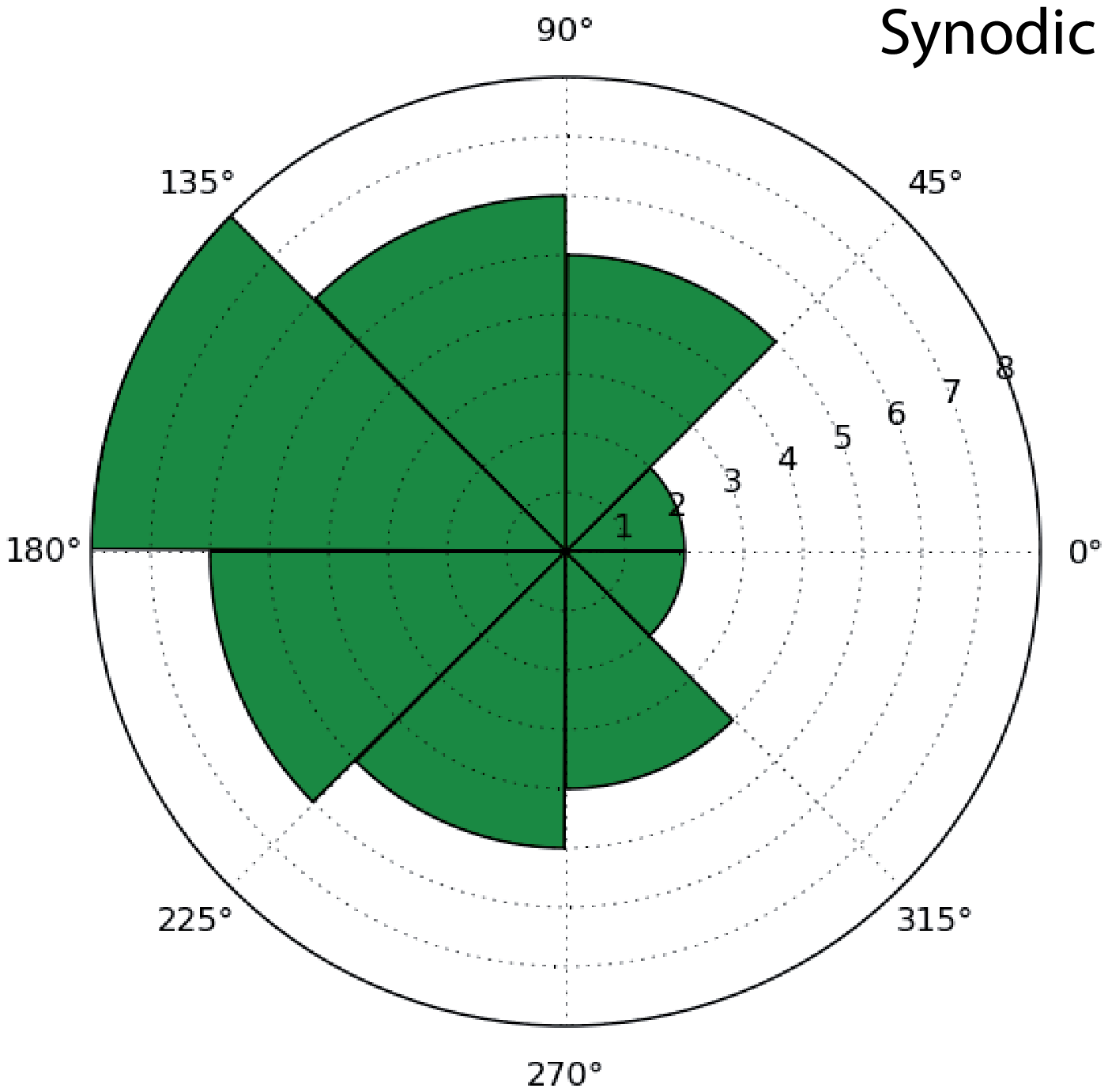}
   \includegraphics[scale=0.5]{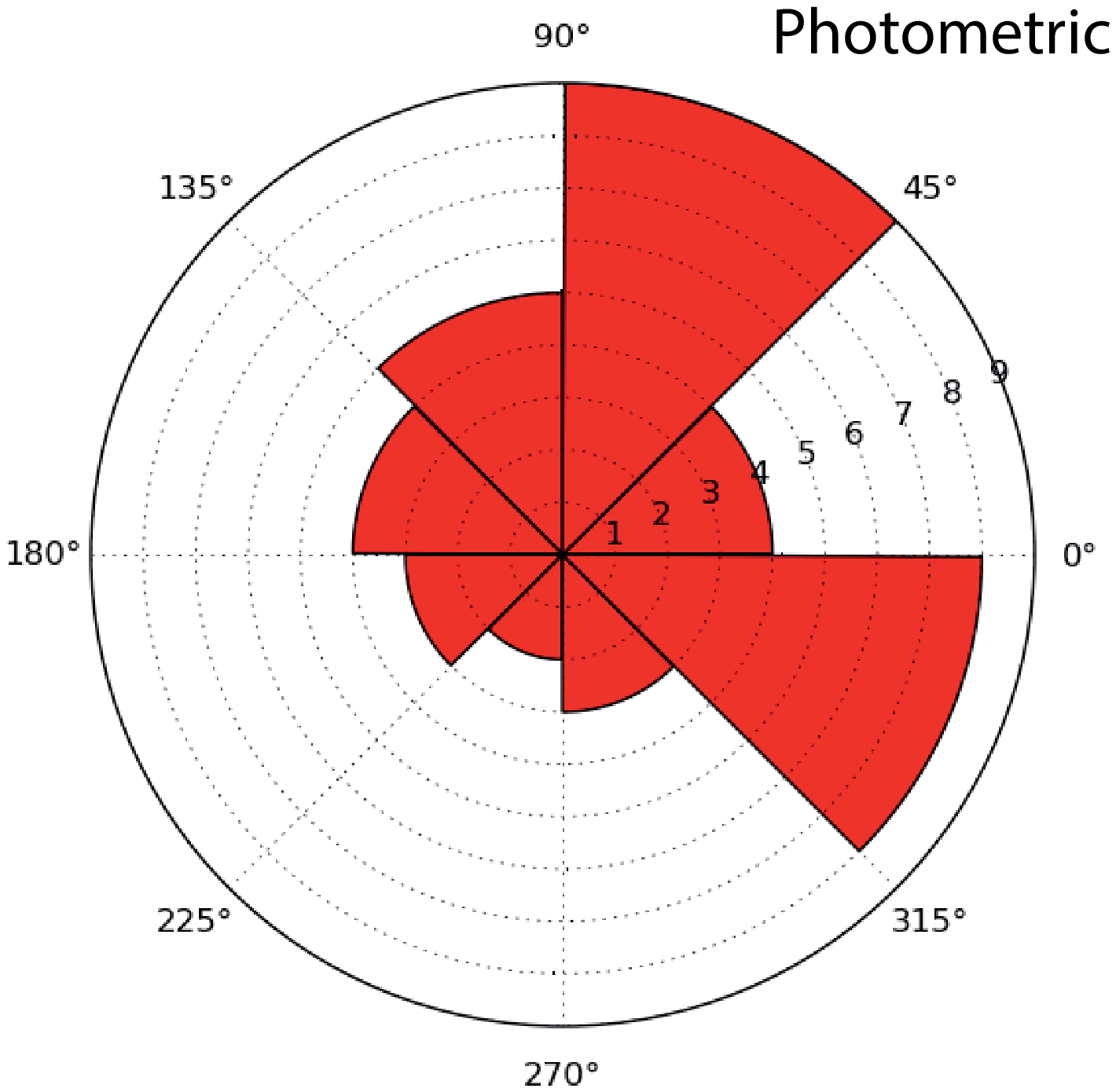}
	\caption{Histogram for flare occurrence in polar coordinates.\textbf {Top}: Phase folded orbital period. \textbf {Middle}: Phase folded synodic period. \textbf{Bottom}: Phase folded photomeric period.}
        \label{fig:histograma}
   \end{figure}

Fig.~\ref{fig:histograma} shows three histograms, in polar coordinates, that represents the number of flare events vs. phase, including all data sets. We plot every polar histogram phase folded with the three main periods found in the system. The histogram for the photometric period suggested that there are privileged areas over stellar surface from which flare precursor loops arise. 
The histogram for the orbital period phase folded suggested a possible day-side effect.

\begin{figure}
  \centering
     \includegraphics[scale=0.48] {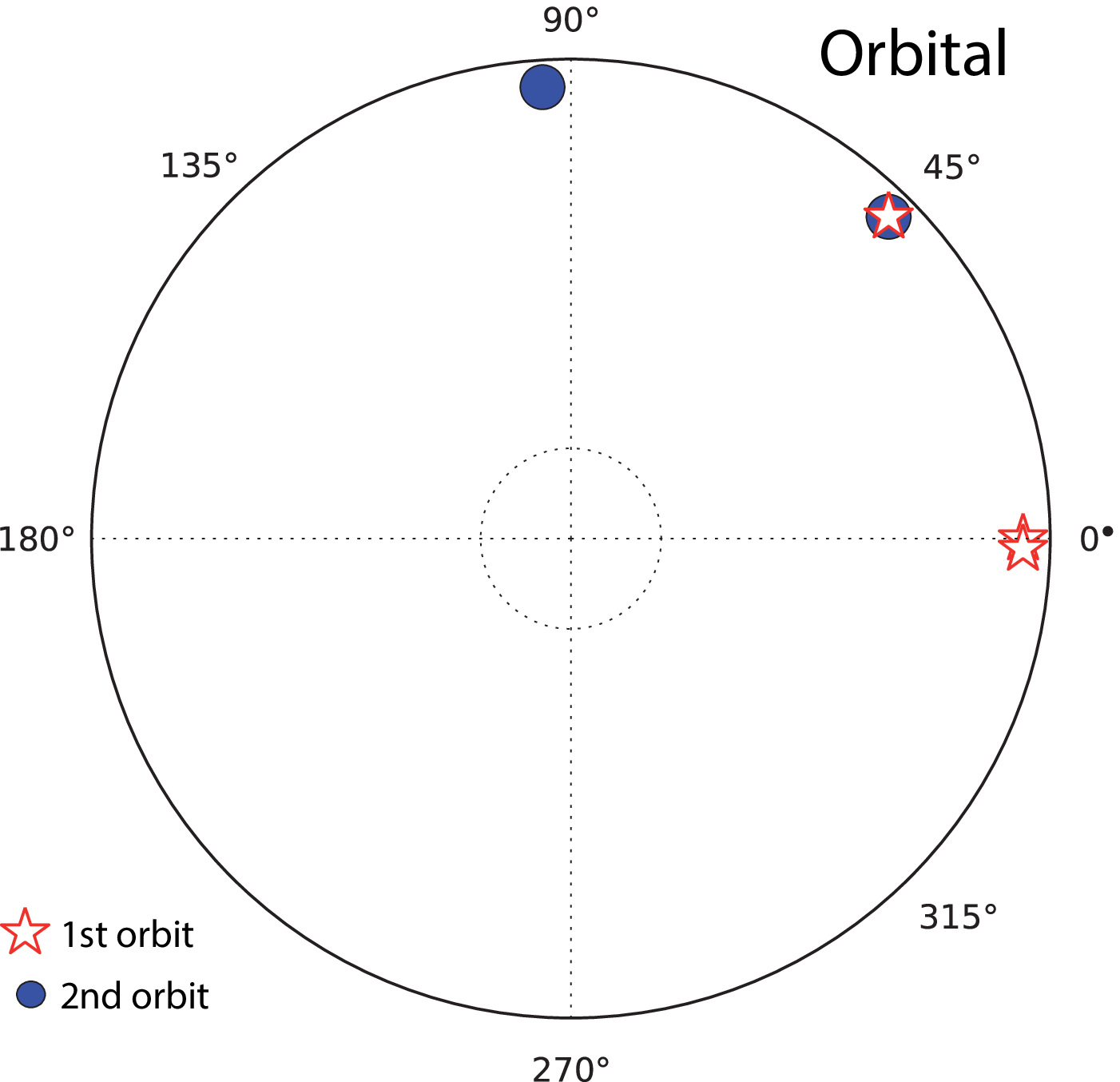}
    \includegraphics[scale=0.48]{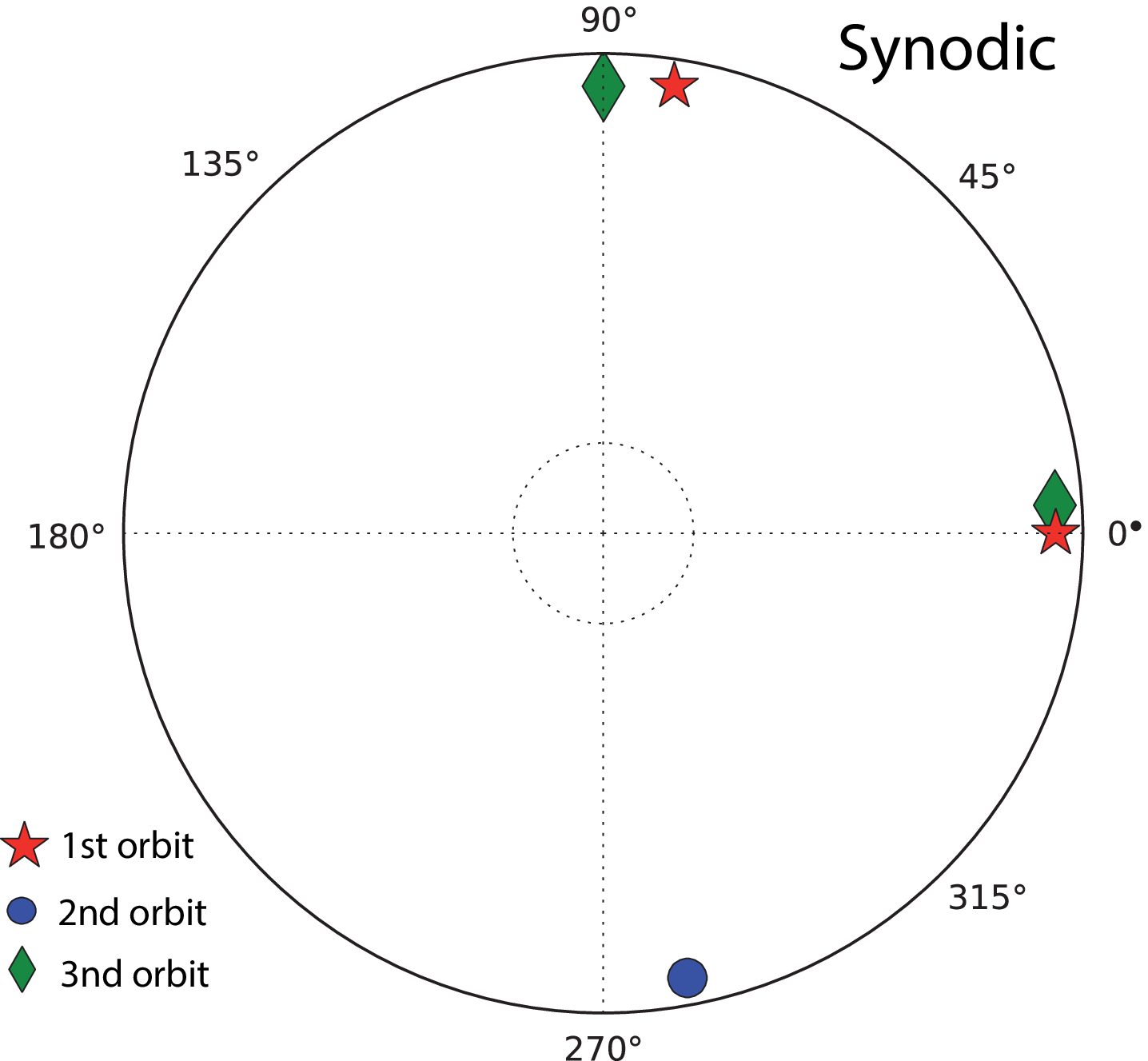}
    \includegraphics[scale=0.48]{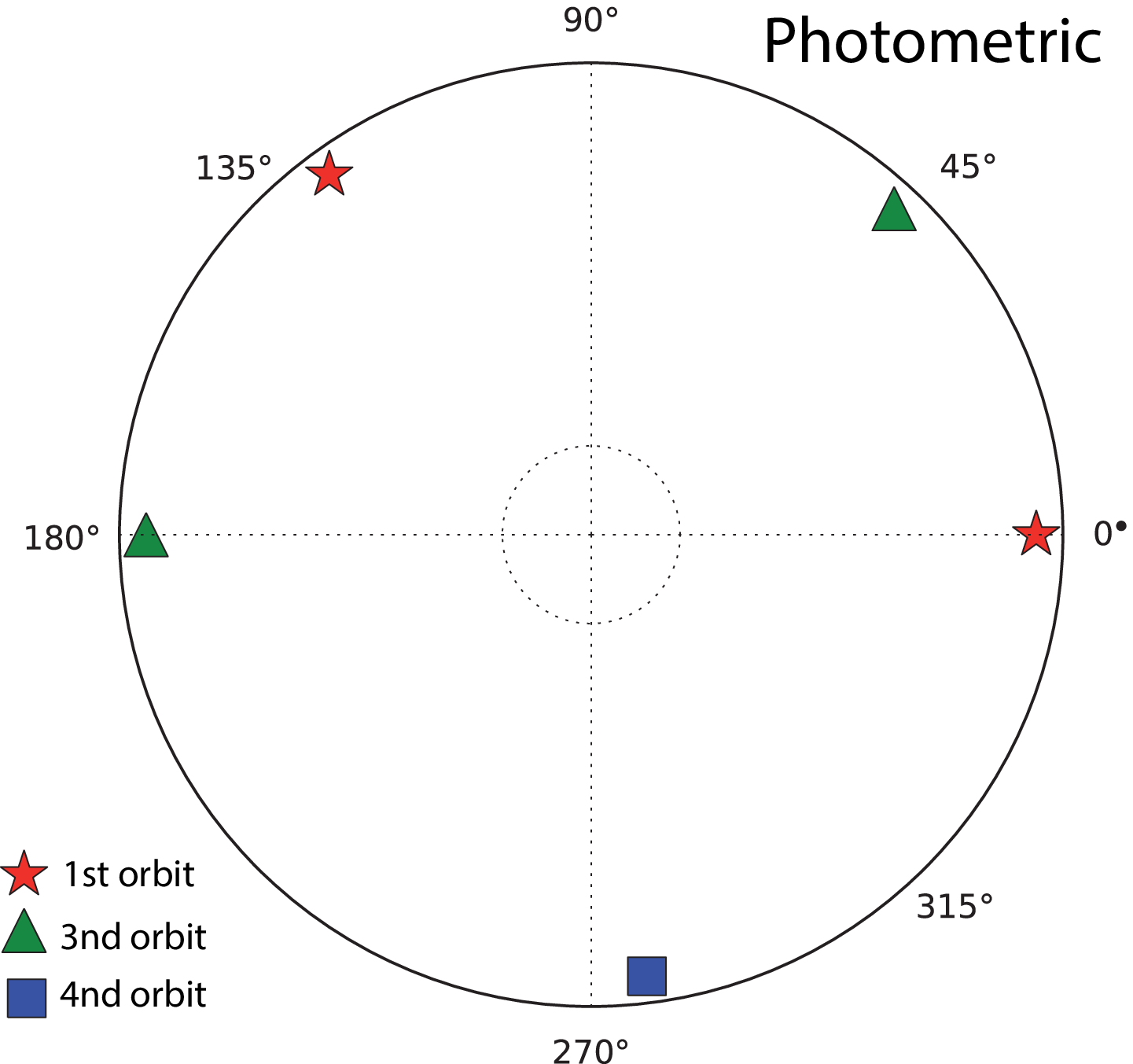}
	\caption{Flare events for the Foces08B run.\textbf {Top}: Phase folded orbital period. \textbf {Middle}: Phase folded synodic period. \textbf{Bottom}:  Phase folded photometric period. The number of orbits for every period are marked with different symbols and colours, that are explained at each figure.}
        \label{fig:vueltas}
   \end{figure}

The fact that we find flares occurring at all phases suggests that they form as stochastic processes. 
We also note that the interaction of the stellar and planetary magnetic fields is not occurring continuously, but at some phases determined by their motions, and also by the variations in the level of the stellar activity between seasons.
To explore in detail, we represent in Fig.~\ref{fig:vueltas} the flare occurrence vs. phase, for the FOCES 08B run, that cover several times the three periods. By folding the data with the orbital period it seems as if the clustering of flares results to be at similar phase at each {\bf orbit}. Also it can be seen that flare clustering appears in multiples of synodic period, suggesting that flare triggering is related to successive passes of the planet over or near the active longitudes.\\

From all of these results, we can hypothesise that there are two possible sources for flares. One related to the geometry of the system and the relative movement of star-planet, that could modulate with synodic period of the system or multiple. And other that is purely stochastic, and is related to the evolution of the magnetic field structures of active regions.

\section{Discussion}

One of the main goals of this study has been to test if the 7.78 day signal was present in all the parameters measured in a same spectra. The simultaneous Bayesian analysis of RV and activity indices reveals that the 7.78 day period is not an artificial signal. This period is not present in activity indices (only in the RV data), whereas we found that activity is modulated by the other two significant periods (photometric and synodic).

Lanza (2011) have suggested that the presence of a close orbiting planet may perturb the stellar magnetic field, and affect the formation of magnetic loops and their evolution and reconnection in the space between the star and the planet. Since magnetic loop reconnection is the invoked physical mechanism to explain flare phenomena, it has a strong stochastic component. In addition to this "geometrical" contribution in flare occurrence (related to the movement of the planet and the star in their orbits), there is a second source of flare events, the magnetic SPI. The successive passages of the planet over the active longitude(s) could trigger the emergence of magnetic flux from the subconvective layers to the upper layers. It can happen that after a number of passages of the planet, the magnetic field intensity reaches the threshold needed for the planet-induced perturbation to be effective.\\

\begin{figure*}

   \includegraphics [scale=0.65] {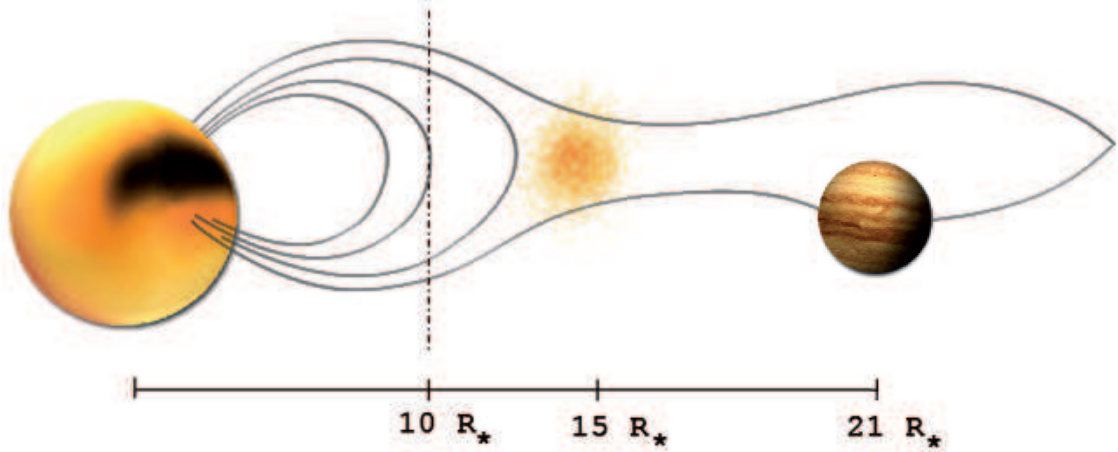}
   \centering
	\caption{Artistic impression (M. Hern\'an-Obispo and S. Recuero) for the BD+20 1790 system, discussed in text. The co-rotation radius is about 10 R$_{*}$, whereas the estimated height for the prominence-like structures is up to 15R$_{*}$. The planet semi-major axis is about to 21R$_{*}$. }
        \label{fig:ilustracion}
   \end{figure*}

In light of the evidence presented in this paper, we hypothesise the following scenario with the aid of an artistic impression (Fig.~\ref{fig:ilustracion}) illustrating the possible geometry of BD +20 1790 and its close orbiting planet. Fig.~\ref{fig:ilustracion} shows a star with a phostosphere containing a large active region. We could interpret our observation of the light curves having no flat regions as evidence that the spots are always in view at high latitude (although we cannot recover the latitude information from our photometry analysis).

Without the velocity information that spectroscopic data can bring, it is not possible to accurately recover latitude information. Helmet streamer-like structures are formed at the top of the X-ray loops anchored to the photospheric active region. These semi-open structures, the helmet streamers, play a key role in the formation of prominences at large heights from the stellar surface and beyond the corotation radius of the star (Ferreira 2000, Jardine \& van Ballegooijen 2005), and could extend up to 20 - 25 R$_{*}$. The prominences lie above the cusp of helmet streamers. The corotation radius for BD+20 1790 is up to 10 R$_{*}$. Due to the inclination of the star (i$\sim $50$^{\rm o}$, Hern\'an-Obispo et al. 2010), the prominence-like structures detected by Hern\'an-Obispo et al. (2005), may be located significantly above the equatorial planet to transit the stellar disc, at high latitude, and at a height of about 15 R$_{*}$ determined following Collier Cameron \& Robinson (1989a). The pronounced double horned peaks that exhibit the H$\alpha$ emission line is a shape typical of emission from circumstellar material at high latitude, that is varying with time (Barnes et al. 2001). Jardine \& van Ballegooijen (2005) suggested that prominence-like structures could form in the stellar wind rather than in the hotter corona, due to the fact they have been detected in some systems with i $<$ 90$^{\rm o}$ (like the widely studied AB Dor, see Cameron \& Robinson 1989 a, b). This is indicating that these prominences lie at regions of high latitude, far away from the equatorial plane, the natural region of equilibrium.
 
One can suppose that a fraction of flares originates in recurrent collisions of helmet-streamers and giant loops that interact with the magnetic field of the planet. This would occur in multiples of synodic period rather than at the orbital period, and also when the planet is near the periastron.

The prominence-like structures are also a source of flare triggering when they became unstable, leading to magnetic reconnection events. Several flaring binary systems have been observed to show prominences (Jeffries 1996, Hall \& Ramsey 1992, 1994).

The orbital phases corresponding to the two measures of magnetic field are 0.81 for the lower value of -19.6 G, and 0.35 for the highest value of -67.8 G (i. e. on nearly opposite sides of the star). 
The rotational phases corresponding for the two measures of magnetic field are 0.12 for the lower value and 0.57 for the higher. Since we could only measure these two points for the magnetic field, we can not infer much about the variation of the field with phase. Future follow up will hep to clarify the field evolution.\\

As we pointed out at Paper I, the presence of a planetary companion is the most plausible hypothesis to explain the overall picture of the BD+20 1790 magnetic field peculiarities. The hypothesis proposed has physical sense, linking the origin of the different phenomena themselves.\\

The completed characterization of the different atmospheric manifestations of stellar activity (photospheric spots, plages and prominences at chromosphere and flares) is beyond the scope of this paper and will be detailed in a forthcoming paper (Hern\'an-Obispo et al., in prep.).

\section{Summary}

In this paper we have presented the simultaneous Bayesian analysis of the RV and activity data for the very active star BD+20 1790. Our results clearly indicate that all the RV data, except CORALIE, contain a strong signal at 7.78 days, compatible with the period of the planetary companion proposed in Paper I. The Bayesian analysis of CORALIE radial velocity data does not show trends of any signal modulating the data, nor the signal proposed by F10, indicating that these data are inconsistent with the rest of the measurements. We conclude that the current evidence supports a planetary companion based on our Bayesian analysis of activity and RV data.

Also, we present a new way of treating the stellar jitter component that contaminates the RV signal. Based on the Bayesian analysis we show that the RV variations actually come from a combination of phenomena, modulated with different periods. Superimposed on the pure planet contribution to the RV curve we found two additional signals: one that varies with the photometric period (2.8 days) and a second that varies with the synodic period (4.36 days).

We present a more accurate orbital solution for the planet after removing the two main contributions of stellar activity to the stellar jitter (associated to photometric and synodic periods).

From the Bayesian analysis of the activity indices we find strong modulation with the synodic period (4.36 days) that leads us to suspect a possible SPI. The presence of a planetary companion also helps us to explain the high rate of flare events in BD+20 1790. We propose two different sources for flare events in this system. The first is related to the geometry of the system and the relative movement of the star and planet, that could modulate with the synodic period of the system or multiple. A second cause of flaring could be purely stochastic and related to the evolution of the stellar surface active regions.

Although we obtained high precision radial velocity data points of BD+20 1790 with HARPS-N spectrograph, matching our result in the RV curve, the comparison with less precise instruments and the lack of coverage of all phases, prevent us using them for further orbital solution acquisition. More data should be added, covering all phases, to obtain a better orbital solution.

This star, as shown by spectropolatrimetric analysis, is a good candidate to perform Zeeman-Dopler Imaging (ZDI), that can allow us to reconstruct the surface magnetic topology. With extrapolation techniques (see i.e. Donati et al. 2008), ZDI gives us access to the 3D magnetic field structure, that could be the most powerful tool to study in detail the magnetic interaction of a close-in massive planet and its host star.\\

The Bayesian tools have proven efficient in detecting low-amplitude signals in RV data (e.g. the independent detection of GJ 163 d based on much smaller data set; compare the results of Tuomi \& Anglada-Escude (2013) and Bonfils et al. (2013)) and in avoiding the detection of false positives (e.g. Gregory 2011; Tuomi 2011). Therefore, with the highly significant detection of the three periodic signals from the combined RV data, we can claim that these signals are not simply artifacts caused by data sampling and noise. Furthermore, all these signals have a good phase coverage (Fig. \ref{fig:mikkofit}), which indicates that the gaps in the data cannot be the cause of the observed periodicities, although the applicability of Bayesian data analysis techniques are not affected by such gaps. Also, while the individual data do not allow the detection of the 2.69 and 4.36 day periodicities in a significant manner, combining them makes the significance of these signals very high. This result, based on the combined data, underlines the importance of combining different data sets in consistency with their respective error sources, as recently discussed in Tuomi et al. (2011).

\begin{acknowledgements}
Based on observations collected at the German-Spanish Astronomical Center, Calar Alto, jointly operated by the Max-Planck-Institut f\"ur Astronomie Heidelberg and the Instituto de Astrof\'isica de Andaluc\'ia (CSIC). Based on observations made with the Italian Telescopio Nazionale Galileo (TNG) operated on the island of La Palma by the Fundaci\'on Galileo Galilei of the INAF (Instituto Nazionale di Astrofisica); and with he Nordic Optical Telescope, operated on the island of La Palma jointly by Denmark, Finland, Iceland, Norway, and Sweden; and with the Mercator Telescope, operated on the island of La Palma by the Flemish Community (obtained with the HERMES spectrograph, which is supported by the Fund for Scientific Research of Flanders (FWO), Belgium , the Research Council of K.U.Leuven, Belgium, the Fonds National Recherches Scientific (FNRS), Belgium, the Royal Observatory of Belgium, the Observatoire de Gen\`eve, Switzerland and the Th\"uringer Landessternwarte Tautenburg, Germany), all operated at the Spanish Observatorio del Roque de los Muchachos of the Instituto de Astrof\'isica de Canarias.
This work was supported by the Spanish Ministerio de Educaci\'on y Ciencia (MEC) under grant
AYA2005-02750, Ministerio de Ciencia e Innovaci\'on (MICINN) under grant
AYA2008-06423-C03-03, AstroMadrid S2009/ESP-1496. M. Hern\'an-Obispo wants to especially thanks to P. Figueira and M. Marmier for their availability to help and
 for providing us with the CORALIE data used in the activity study
 and the information requested. M. C. G\'alvez-Ortiz acknowledges financial support by the Spanish MICINN under
 the Consolider-Ingenio 2010 Program grant CSD2006-00070: First Science with
 the GTC (http://www.iac.es/consolider-ingenio-gtc). She also acknowledges the
 support of a JAE-Doc CSIC fellowship co-funded with the European Social Fund
 under the program {\em 'Junta para la Ampliaci\'on de Estudios'}. Financial support was also provided by the Spanish Ministerio de Ciencia e Innovaci\'on and Ministerio de Econom\'ia y Competitividad under AyA2011- 30147-C03-03 grant.
 M. Tuomi, J. R. Barnes, M. C. G\'alvez-Ortiz and D. J. Pinfield acknowledge support from RoPACS during this research, a Marie Curie Initial Training Network funded by the European Commission's Seventh Framework Programme. J. S. Jenkins acknowledges funding from CATA (PB06, Conicyt). We want also to thanks Saskia Prins for her help in MERCATOR observations and HERMES pipeline members (L. Dumortier, Y. Fremat, H. Hensberge, A. Jorissen, S. Van Eck, H. Van Wickel) for their help in the data reduction. M. Hern\'an-Obispo thanks Sito Recuero for his help improving the Fig. 18.
This research has made use of the SIMBAD database, operated at CDS, Strasbourg, France. 
The authors are grateful to an anonymous referee for valuable comments and suggestions which have helped improve this paper.
\end{acknowledgements}

\end{document}